\documentclass[a4paper,11pt]{article}
\pdfoutput=1
\usepackage{jheppub}
\usepackage[T1]{fontenc}
\usepackage{amsmath,amssymb,euscript}
\usepackage{mathrsfs}
\usepackage{braket}
\usepackage{slashed}
\usepackage{xspace}
\usepackage{color}
\usepackage{hyperref}
\usepackage{epsfig}
\usepackage[table]{xcolor}
\usepackage{xspace}
\usepackage{verbatim}
\usepackage{multirow}
\usepackage{booktabs,graphicx}
\usepackage{mathtools}
\usepackage{tabulary}
\usepackage{soul}
\usepackage{caption}
\usepackage{subcaption}
\usepackage{orcidlink}
\usepackage{xfrac}
\usepackage{ragged2e}
\usepackage{longtable}
\usepackage{float}
\usepackage{doi}
\usepackage{placeins}

\title{\boldmath Hunting for Long-Lived ALPs in Top Events via Displaced Dielectron Pairs at the LHC}

\author[a]{Juliette Alimena \orcidlink{0000-0001-6030-3191},}
\author[a,b]{Freya Blekman \orcidlink{0000-0002-7366-7098},}
\author[a]{Cesare Cazzaniga \orcidlink{0000-0003-0001-7657},}
\author[c]{Xabier Cid Vidal \orcidlink{0000-0002-0468-541X},}
\author[d]{Jeremi Niedziela \orcidlink{0000-0002-9514-0799},}
\author[a,b]{Lovisa Rygaard \orcidlink{0000-0003-3192-1622},}
\author[c]{Rosal\'ia Soto Rey \orcidlink{0009-0002-8804-9476}}

\affiliation[a]{Deutsches Elektronen-Synchrotron DESY, Notkestr. 85, 22607 Hamburg, Germany}
\affiliation[b]{Institut f\"ur Experimentalphysik, Universit\"at Hamburg, Luruper Chaussee 149, 22761 Hamburg, Germany}
\affiliation[c]{Instituto Galego de Física de Altas Enerxías (IGFAE), Universidade de Santiago de Compostela, Santiago de Compostela, Spain}
\affiliation[d]{Interuniversity Institute for High Energies (IIHE), Vrije Universiteit Brussel, Pleinlaan 2, 1050 Brussels, Belgium}

\emailAdd{rosalia.soto.rey@cern.ch}

\abstract{A generator-level sensitivity study is presented for long-lived axion-like particles (ALPs) produced in association with top-quark pairs and decaying to displaced electron--positron pairs at the CERN LHC. The prompt objects from the top quark--antiquark system provide independent handles for event selection, allowing the event to be selected independently of the displaced electron pair and extending the search to ALP masses below the dimuon threshold. A search strategy motivated by the CMS detector is developed for ALP masses between 17~MeV and 10~GeV and proper decay lengths between 0.1~mm and 1~m. Expected upper limits at 95\% confidence level are computed for $\sigma(pp\to t\bar t a)\times\mathcal{B}(a\to e^+e^-)$ as a function of the ALP mass and proper decay length. In the most sensitive region of the parameter space considered, values of the cross section times branching fraction of order $10^{-3}$~pb can be probed for an integrated luminosity of 500~fb$^{-1}$ at $\sqrt{s}=13$~TeV. These projections do not include detector-level displaced-electron reconstruction or instrumental backgrounds.
}

\begin{document} 
\maketitle
\flushbottom

\section{Introduction}
\label{sec:intro}

Long-lived particles (LLPs) arise in many extensions of the Standard Model (SM)~\cite{lee2019collider, alimena2020searching, knapen2023guide}. Macroscopic lifetimes can result, for example, from small couplings to SM particles, approximate symmetries, limited decay phase space, or interactions mediated by a high mass scale. Collider searches for displaced decays therefore probe regions of parameter space that can be difficult to access with searches for promptly decaying or detector-stable particles. The distinctive spatial structure of displaced decays can also provide strong discrimination from SM processes, making LLP searches an important part of the broader search program for physics beyond the SM.

Exploiting this sensitivity experimentally can nevertheless be challenging, particularly when the visible decay products are soft, displaced, or otherwise difficult to trigger on directly. One approach is to consider LLP production in association with prompt particles that provide independent handles for event selection. Top-quark pair production is especially useful in this respect: the top-quark decays provide energetic leptons, jets, $b$ jets, and missing transverse momentum that can be used to select the event without relying entirely on the displaced LLP decay products. 

Light axion-like particles (ALPs), whose decay products are soft and difficult to trigger on directly, are a natural target for this strategy, which was investigated in Ref.~\cite{rygaard2023} for long-lived ALPs produced in association with a top quark–antiquark pair and decaying to displaced dimuons. In ALP scenarios with comparable dimensionless fermion couplings, the effective pseudoscalar interaction is proportional to the fermion mass, making top-associated production particularly well motivated. The strategy is, however, not specific to ALPs: it may also be relevant for other light neutral long-lived states produced in association with a $t\bar t$ pair and decaying to displaced leptons. Related displaced signatures arise, for example, in scenarios with dark photons arising from a spontaneously broken $U(1)_D$ with kinetic mixing, whose couplings to charged fermions are governed by electric charge rather than fermion mass, as well as long-lived scalars mixing with the Higgs boson, heavy neutral leptons, and light states from confining hidden sectors~\cite{Strassler:2006im,Curtin:2013fra}. In what follows, we adopt the ALP as a concrete benchmark; quantitative interpretations in other models would require accounting for differences in production and decay kinematics.

In this work, we extend the top-associated strategy to the process $pp \to t\bar t a, a\to e^+e^-$. The dielectron channel opens access to ALP masses below the dimuon threshold, which are inaccessible to the displaced-dimuon strategy of Ref.~\cite{rygaard2023}, since ALPs with $m_a < 2m_\mu$ cannot decay to dimuons. This extension is experimentally nontrivial: displaced-electron reconstruction poses additional challenges relative to muon reconstruction~\cite{CMS:2021kdm,CMS:2026acz}, and at low ALP masses the electrons can be both soft and highly collimated. This combination of displacement, low transverse momentum, and collimation lies outside the regime for which standard electron triggering and reconstruction are optimized. Displaced-dielectron searches at ATLAS and CMS have often used photon triggers with relatively high $p_T$ thresholds~\cite{ATLAS:2026riv,CMS:2021kdm}, while more recent strategies have begun to exploit dedicated displaced-track or low-$p_T$ electron reconstruction techniques~\cite{ATLAS:2024vnc,CMS:2026acz}.  The associated $t\bar t$ system instead provides trigger and event-selection handles that are independent of the displaced electron pair, enabling this lower-mass regime to be probed without relying on the displaced electrons themselves for triggering or event selection.

At the LHC, direct searches for long-lived ALPs decaying to electron pairs have so far received little attention, and, to our knowledge, the signature of a long-lived ALP produced in association with a $t\bar t$ pair and decaying to an electron pair has not yet been studied, either phenomenologically or experimentally. As a first step towards an experimental search, it is therefore necessary to establish the expected sensitivity of this channel and identify the key handles for event selection. Here we perform a generator-level sensitivity study for long-lived ALPs with masses between 17 MeV and 10 GeV and proper decay lengths between 0.1 mm and 1 m, using the CMS detector geometry and performance as guidance for the search strategy. For proper decay lengths beyond the range considered, the sensitivity of this tracker-based strategy is expected to decrease as an increasing fraction of ALPs decay outside the tracking volume, with the precise decay-position distribution depending on the ALP boost. The lower edge of the mass range also covers the mass scale associated with the X17 anomaly, where excesses in $e^+e^-$ pair production in nuclear transitions of $^{8}$Be, $^{4}$He, and $^{12}$C have been interpreted as a new light boson with a mass of about 17 MeV~\cite{Krasznahorkay:2015iga, Krasznahorkay:2021joi, Krasznahorkay:2019lyl}, with pseudoscalar, ALP-like interpretations among those considered~\cite{Ellwanger:2016wfe}. More generally, the range is chosen to extend the reach of CMS toward the low-mass region.

The primary results of the study are presented as expected upper limits on $\sigma(pp\to t\bar t a) \times \mathcal B(a\to e^+e^-)$ as a function of the ALP mass and proper decay length. The study is intended as an idealized estimate of the potential of the top-associated displaced-dielectron signature; detector-level displaced-electron reconstruction and instrumental backgrounds are outside its scope. We additionally interpret the results in the top scenario of Ref.~\cite{rygaard2023}, in which only the ALP coupling to top quarks is nonzero at the cutoff scale.

The remainder of this paper is organized as follows. The search strategy is introduced in Sec.~\ref{sec:strategy}, and the simulated samples and event selection are described in Sec.~\ref{sec:sim_sel}. The projected sensitivity is presented in Sec.~\ref{sec:sensitivity}, and a summary is given in Sec.~\ref{sec:summary}.

\section{Search strategy} \label{sec:strategy}

\begin{figure}[h]
    \centering
    \begin{minipage}[c]{0.47\linewidth}
        \centering
        \includegraphics[width=\linewidth]{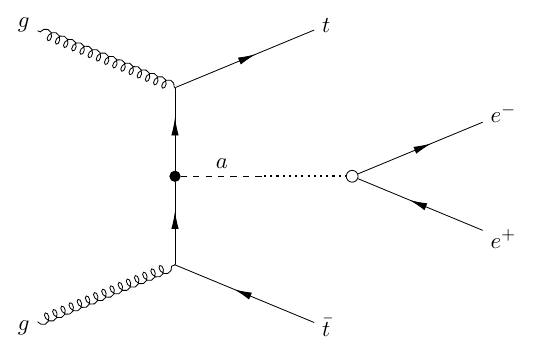}
    \end{minipage}
    \hfill
    \begin{minipage}[c]{0.47\linewidth}
        \centering
        \includegraphics[width=\linewidth]{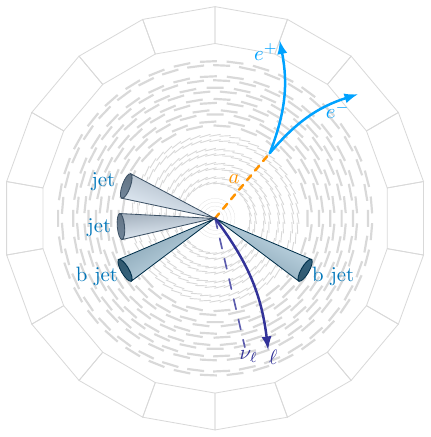}
    \end{minipage}
    \caption{Left: Representative Feynman diagram of the LHC process $pp \to t\bar{t}a$, $a\to  e^+ e^-$. The black circle indicates the effective ALP-top coupling, and the white circle indicates the effective ALP-electron coupling. In the top scenario considered in Appendix~\ref{sec:top-scenario}, the latter is radiatively induced by the ALP-top coupling. The dotted line indicates that the ALP decay is displaced. Right: Sketch of the signal event topology.}
    \label{fig:diagram}
\end{figure}

We consider the process $pp \to t\bar t a, a\to e^+e^-$ for ALP masses above the dielectron threshold, $m_a\geq 2m_e$. A representative production diagram is shown in Fig.~\ref{fig:diagram} (left). The experimental signature consists of a prompt $t\bar t$ system together with an electron--positron pair originating from a displaced ALP decay, as illustrated in Fig.~\ref{fig:diagram} (right). Triggering on and reconstructing $t\bar t$ events is well established at the LHC; leptons, jets, $b$ jets, and missing transverse momentum from the top-quark decays can provide trigger and event-selection handles independently of the displaced-dielectron system. In this study, we include all $t\bar t$ decay modes, and assume that the events can be selected with 100\% efficiency using the $t\bar t$ system. Since the achievable efficiency depends on the decay channels and selection criteria used, the impact of a 10\% efficiency is discussed in Sec.~\ref{sec:sensitivity}.

The search targets ALP decays within the CMS tracking volume, i.e., inside the outer radius of the tracker of about 1.1~m~\cite{hayrapetyan2024development}, where the two charged decay products can in principle be reconstructed as displaced tracks and associated with a common secondary vertex. Standard CMS electron reconstruction combines tracks reconstructed in the silicon tracker with energy deposits in the electromagnetic calorimeter and is optimized primarily for electrons produced close to the interaction point~\cite{CMS:2020uim}. Displaced electrons may therefore fail track-seeding or identification requirements, particularly for decays occurring beyond the innermost tracker layers~\cite{CMS:2026acz}.

The analysis is intended as an idealized generator-level sensitivity estimate rather than a complete detector-level projection. No explicit displaced-electron reconstruction efficiency is applied. Although displaced-electron reconstruction efficiencies have been reported for dedicated CMS algorithms~\cite{CMS:2026acz}, they are derived for different signal topologies and kinematic regimes and are not directly transferable to the highly collimated dielectron signatures considered here. When constructing trajectory-dependent observables, in particular the dielectron distance of closest approach and the secondary-vertex position entering the collinearity angle defined in Sec.~\ref{sec:signal-selection}, the electron trajectories are approximated as straight lines from the generator-level production points along the corresponding momentum directions, neglecting their curvature in the CMS magnetic field. This approximation is most consequential for low-momentum electrons, for which the curvature can be significant. The physics background is modeled using simulated inclusive $t\bar t$ events, which are expected to dominate the sample of events with a $t\bar t$-like topology and displaced $e^+e^-$ pairs, for example from heavy-flavor hadron decays. Other physics processes, such as single-top production or $t\bar t$ production in association with a single boson, are expected to be subdominant and are not considered in this phenomenological study. Backgrounds arising from additional inelastic proton-proton interactions within the same or neighboring bunch crossings (``pileup''), detector-material interactions, object misreconstruction, and other instrumental effects are not included. The resulting projections should therefore be interpreted within these assumptions.

\begin{figure}[h]
    \centering
    \includegraphics[width=0.9\linewidth]{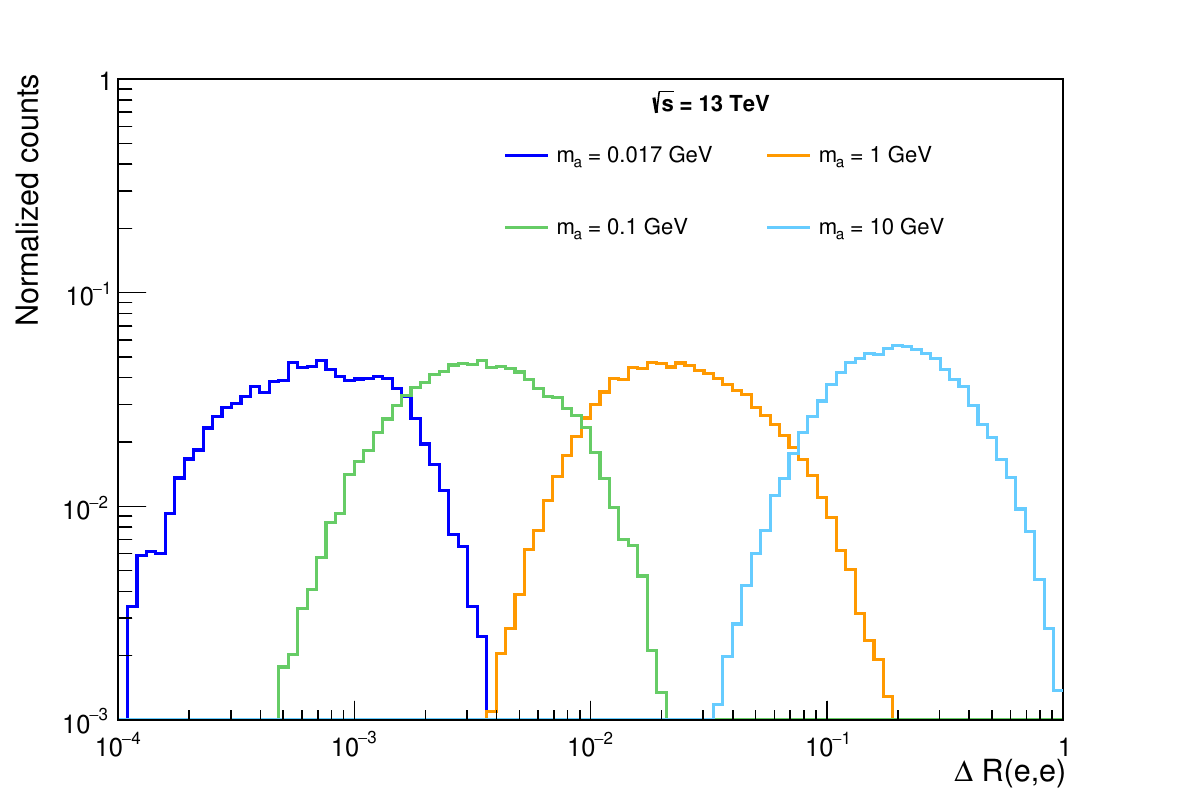}
    \caption{Distribution of the angular separation $\Delta R(e,e)$ between the electron and positron after the full signal selection defined in Sec.~\ref{sec:signal-selection}, showing several signal benchmarks with $c\tau = 1 \,\, \mathrm{mm}$. The distributions are normalized to unit area.}
    \label{fig:deltaR}
\end{figure}

An additional experimental challenge arises at low ALP masses, where the electron and positron can be highly collimated. Related topologies have been targeted in searches for light long-lived particles decaying to collimated leptons or light hadrons~\cite{ATLAS:2022izj} and, more recently, for highly collimated electron--positron pairs from light pseudoscalar decays~\cite{CMS:2025zwc}. As shown in Fig.~\ref{fig:deltaR}, the characteristic angular separation decreases rapidly with decreasing $m_a$, reaching values well below $\Delta R(e,e)=0.01$ for the lowest masses considered. The highly boosted low-mass ALPs can also decay at large transverse displacements, reducing the tracker lever arm available after the decay. These effects can make it difficult to resolve two distinct electron trajectories and reconstruct two separate electron objects. At sufficiently small separations, overlapping electromagnetic showers can also complicate the reconstruction in the calorimeter. The efficiency for reconstructing such highly collimated, displaced electron pairs is not modeled in the present generator-level study. However, several features of the CMS reconstruction may mitigate some of these effects. First, the electron track information relevant for vertexing is obtained from fits to the tracker hits rather than from the calorimeter shower position. Second, the straight-line treatment considered in this study neglects the opposite bending of the electron and positron trajectories in the CMS magnetic field, which can increase their separation as they propagate through the tracker.

\section{Simulation and event selection}
\label{sec:sim_sel}
The projected sensitivity is evaluated using simulated samples of the signal process and of inclusive SM $t\bar t$ production, which is used to model the physics background in this study. The event simulation and normalization are described in Sec.~\ref{sec:simulation}, while the displaced-dielectron event selection is presented in Sec.~\ref{sec:selection}.

\subsection{Event simulation}\label{sec:simulation}
Hard-scattering events for the signal and background processes are generated using \texttt{MadGraph5\_\allowbreak aMC@NLO} v. 3.4.2~\cite{alwall2014automated} for proton--proton collisions at $\sqrt{s} = 13 \,\, \mathrm{TeV}$. The background sample is generated using the NNPDF3.1 NLO PDF set~\cite{Ball:2017nwa}, while the signal samples use the MadGraph default NNPDF2.3 NLO set~\cite{Ball:2012cx}. Parton showering and hadronization are simulated using \texttt{PYTHIA8} v. 8.316~\cite{bierlich2022comprehensive} with the Monash 2013 tune~\cite{Skands:2014pea}. The hard-scattering processes are generated at leading order in QCD, and the top-quark mass is set to $m_t = 172 \,\, \mathrm{GeV}$.

The simulated background sample contains approximately $4 \times 10^7$ generated inclusive $t\bar{t}$ events. These events are subsequently normalized to the expected yield using the measured $t\bar{t}$ production cross section at $\sqrt{s}=13\,\, \mathrm{TeV}$, $\sigma(p p \to t\bar{t}) = 888 \,\, \mathrm{pb}$~\cite{cms2017measurement}.

Signal events are generated using the UFO implementation of the ALP effective theory employed in Ref.~\cite{rygaard2023}. Samples are generated for several benchmark ALP masses,
\begin{equation*}
m_a \in \{0.017, \,\, 0.1, \,\, 0.2, \,\, 0.3, \,\, 0.5, \,\, 1, \,\, 2, \,\, 5, \,\, 7, \,\, 10\} \,\, \mathrm{GeV},
\end{equation*}
and proper decay lengths,
\begin{equation*}
c\tau \in \{10^{-1}, \,\, 10^0, \,\, 10^1, \,\, 10^2, \,\, 10^3
\} \,\, \mathrm{mm}.
\end{equation*}
The signal normalization is treated as a free parameter: the primary results are expressed as limits on $\sigma(pp \to t \bar{t} a) \times \mathcal{B}(a \to e^+ e^-)$ as a function of the ALP mass and its proper decay length.

At the generator level, the ALP is required to have a transverse momentum $p_T~>~20$~GeV and the electrons from the ALP decay are required to have a pseudorapidity of $|\eta| < 3.5$ to reduce the computational cost of the simulation. The quoted signal cross sections and corresponding limits are defined within this generated fiducial phase space. A total of $2 \times 10^5$ events are generated for each ($m_a, c\tau$) signal benchmark, except for the $c\tau = 1000 \,\, \mathrm{mm}$ samples, for which $4 \times 10^5$ events are generated because of the lower selection efficiency.

\subsection{Event selection}\label{sec:selection}
The event selection proceeds in two stages. An initial pre-selection suppresses prompt backgrounds and retains events containing at least one displaced $e^+e^-$ candidate satisfying the requirements listed in Sec.~\ref{sec:preselection}. A subsequent signal selection, described in Sec.~\ref{sec:signal-selection}, further suppresses the $t\bar{t}$ background using the properties of the dielectron pair.

Dielectron candidates in the $t\bar{t}$ background are classified at generator level as resonant or non-resonant, according to whether the electron and positron originate from the same parent particle. The two components are shown separately in the distributions below. This classification is used only to illustrate the composition of the simulated background and does not enter the event selection. Resonant candidates are more signal-like, since the two electrons share a common production point, and are therefore harder to separate from the signal than non-resonant candidates using vertexing information alone. Some prominent resonant contributions can nevertheless be reduced with dedicated mass vetoes.

\subsubsection{Pre-selection} 
\label{sec:preselection}
Each candidate electron is required to be a final-state particle\footnote{Final-state electrons are used to avoid multiple status copies associated with QED final-state radiation. Photons radiated from the electrons are not recombined with the electron momenta, resulting in a finite width of the dielectron mass distribution. In CMS reconstruction, nearby radiated energy can be partially recovered through the superclustering procedure used for electrons~\cite{CMS:2020uim}.} with opposite electric charge to the other electron in the pair and to satisfy $|\eta| \leq 2.5$ and $500 \,\, \mu\mathrm{m} < l_{xy} < 0.8 \,\, \mathrm{m}$, where the transverse displacement $l_{xy} = \sqrt{x^2 + y^2}$ is computed separately for each electron from its generator-level production coordinates, relative to the interaction point. In signal events, both electrons are additionally required at generator level to originate from the ALP decay. An event passes the pre-selection if it contains at least one electron--positron pair for which both electrons satisfy these requirements.

The minimum $l_{xy}$ requirement defines a clearly displaced region and is chosen conservatively relative to the typical transverse impact parameter resolution of CMS tracks~\cite{cms2014description,CMS:2021kdm}. The maximum $l_{xy}$ requirement is motivated by the dimensions of the CMS tracker and is chosen such that the electrons traverse at least three outer-tracker layers after the ALP decay~\cite{hayrapetyan2024development}.

\begin{figure}[h]
    \centering
    \includegraphics[width=0.61\linewidth, trim = 0 70pt 0 170pt, clip]{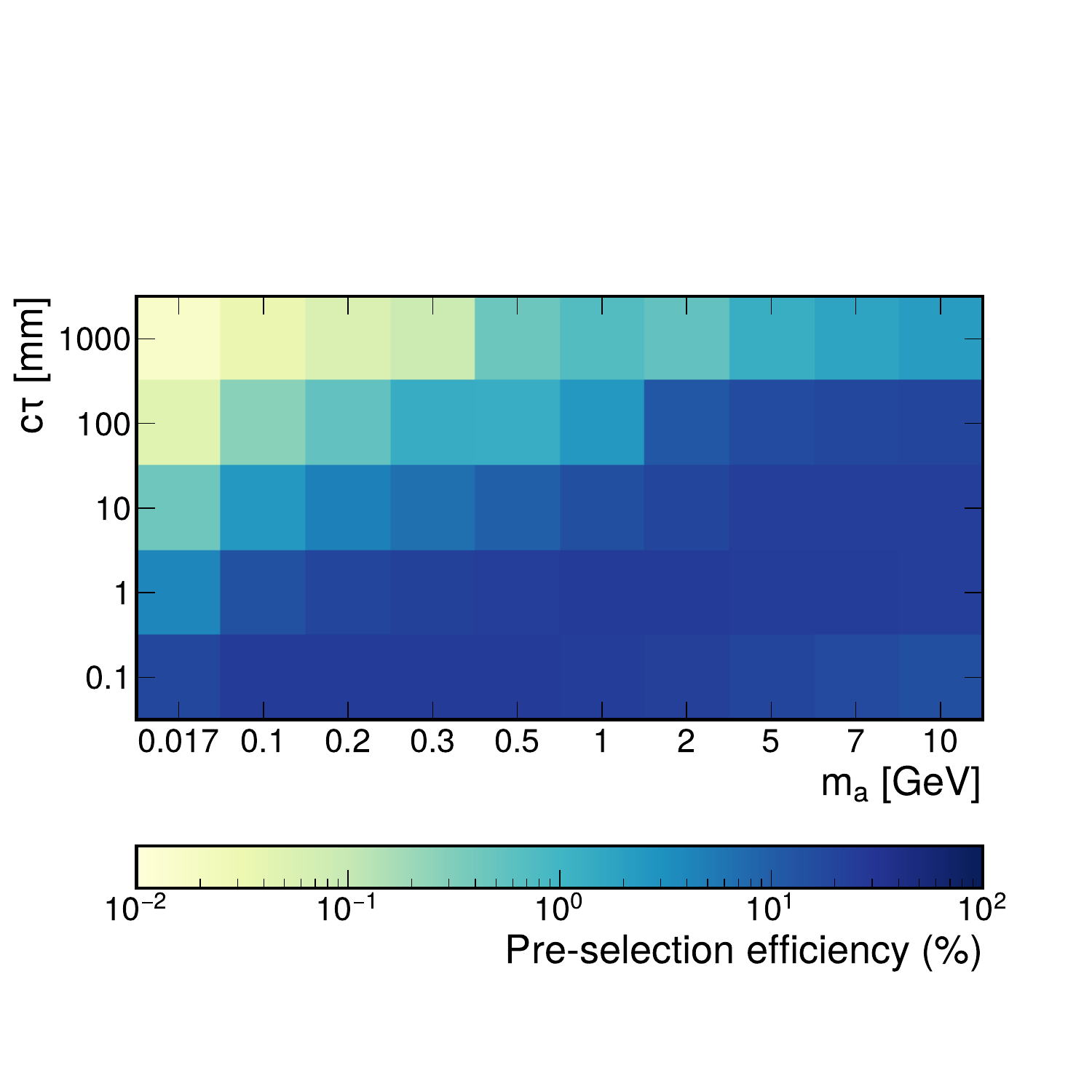}
    \caption{Pre-selection efficiencies for each signal benchmark.}
    \label{fig:eff_preselection}
\end{figure}

The background pre-selection efficiency is found to be $(9.019 \pm 0.005)\%$. The pre-selection efficiencies obtained for each signal benchmark are shown in Fig.~\ref{fig:eff_preselection} and in Appendix~\ref{sec:selection_efficiencies}. At low masses, the efficiency decreases at larger proper decay lengths because an increasing fraction of the ALPs decay beyond $l_{xy} = 0.8 \,\, \mathrm{m}$. At higher masses, the efficiency peaks at intermediate proper decay lengths; for short proper decay lengths and higher masses, the smaller Lorentz boost relative to the lower mass benchmarks causes more ALPs to decay before reaching $l_{xy} = 500 \,\, \mu\mathrm{m}$. The low pre-selection efficiencies for some signal benchmarks therefore primarily reflect the probability for the ALP to decay within the targeted transverse-displacement range rather than the discriminating power of the pre-selection against background. A common pre-selection is retained across the signal benchmarks to define a uniform displaced fiducial region, while the subsequent signal selection provides the main background rejection.

\subsubsection{Signal selection}\label{sec:signal-selection}
The signal selection exploits the distinctive features of the signal and background processes to suppress background contamination and enhance sensitivity to the signal. If more than one dielectron candidate satisfies the pre-selection, the candidate with the smallest value of
\begin{equation*}
    R_{lxy} = \dfrac{\sqrt{(x_{e^-} - x_{e^+})^2 + (y_{e^-} - y_{e^+})^2 }}{\sqrt{(|x_{e^-}| + |x_{e^+}|)^2 + (|y_{e^-}| + |y_{e^+}|)^2 }}
\end{equation*}
is selected, where $(x_{e^\pm}, y_{e^\pm})$ are the transverse coordinates of the production points of the electron and positron. This criterion favors electron--positron pairs whose production points are close to one another, relative to their displacement from the beamline, and is intended as a generator-level proxy for compatibility with a common displaced vertex. The $R_{lxy}$ variable itself is not an experimentally reconstructed observable; in an experimental analysis, the corresponding candidate choice would instead be based on reconstructed track and vertex information.

\begin{figure}[h]
    \centering
    \includegraphics[width=0.9\linewidth]{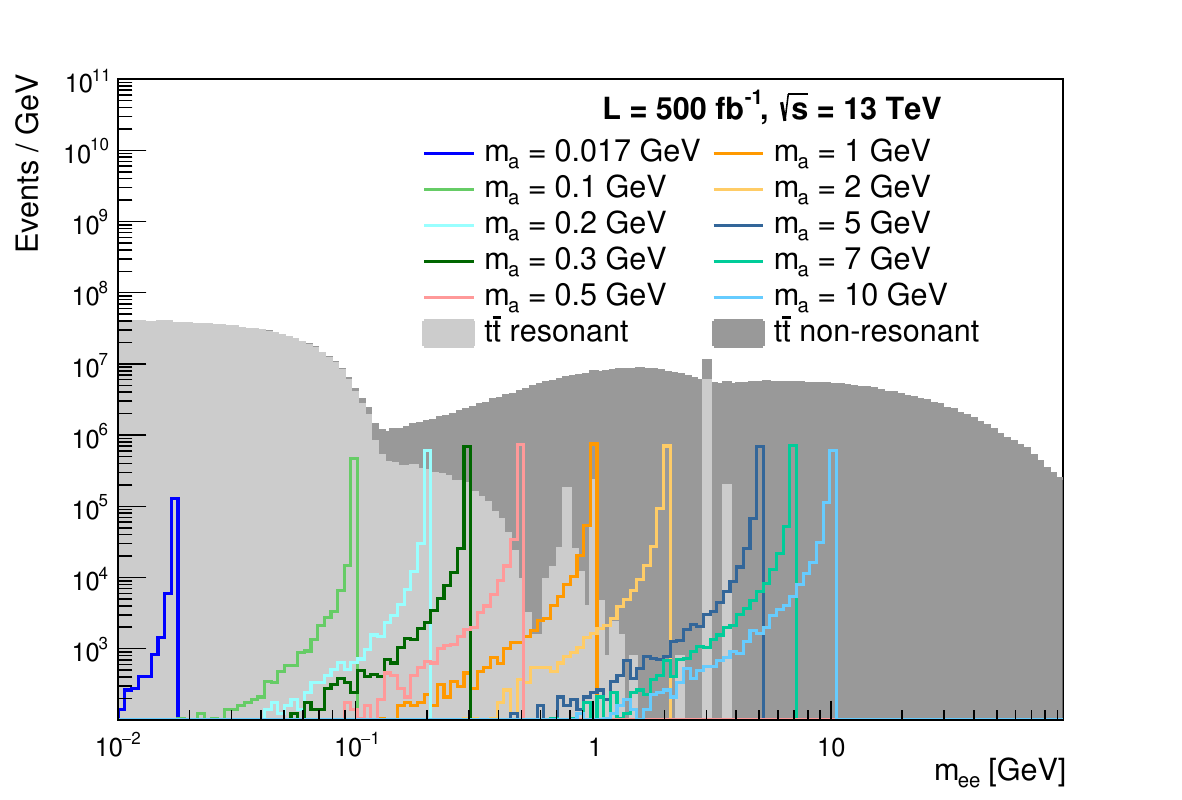}
    \caption{Invariant-mass distribution of the dielectron pair after the pre-selection, showing several signal benchmarks with $c\tau = 1 \,\, \mathrm{mm}$, together with the resonant (light gray) and non-resonant (dark gray) $t\bar{t}$ backgrounds. The background yield is normalized to its experimental cross section, while the signal yields are normalized to an illustrative cross section of  $0.1 \,\, \mathrm{pb}$. An integrated luminosity of $500 \,\, \mathrm{fb^{-1}}$ is assumed.}
    \label{fig:dielectron_mass}
\end{figure}

Fig.~\ref{fig:dielectron_mass} shows the dielectron invariant-mass distribution after the pre-selection. Background contributions from the $J/\psi$, $\psi(2S)$, and $Z$ boson resonances are reduced by rejecting candidates whose invariant masses $m_{ee}$ fall within the windows listed in Table~\ref{tab:mass_vetoes}.

\begin{table}[h]
\begin{center}
\begin{tabular}{c|c|}
\cline{2-2}
 & Veto window {[}GeV{]} \\ \hline
\multicolumn{1}{|c|}{$J/\psi$} & $3.05 < m_{ee} < 3.13$ \\ \hline
\multicolumn{1}{|c|}{$\psi(2S)$} & $3.50 < m_{ee} < 3.86$ \\ \hline
\multicolumn{1}{|c|}{$Z$} & $86.63 < m_{ee} < 95.75$ \\ \hline
\end{tabular}
\end{center}
\caption{Excluded invariant-mass regions.}
\label{tab:mass_vetoes}
\end{table}

\begin{figure}[!htbp]
    \centering
    \includegraphics[width=0.47\linewidth,
        trim={15pt 0 5pt 0},
        clip]{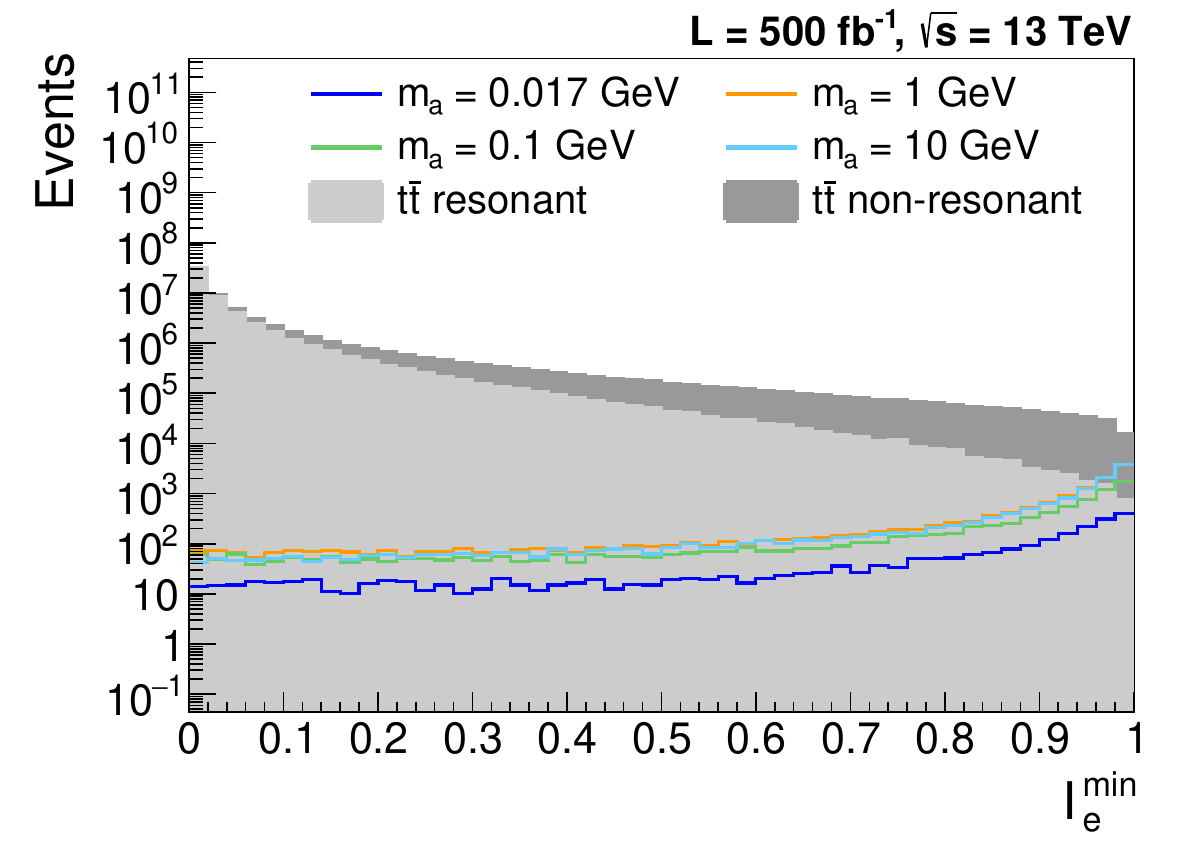}
    \hfill
    \includegraphics[width=0.47\linewidth,
        trim={15pt 0 5pt 0},
        clip]{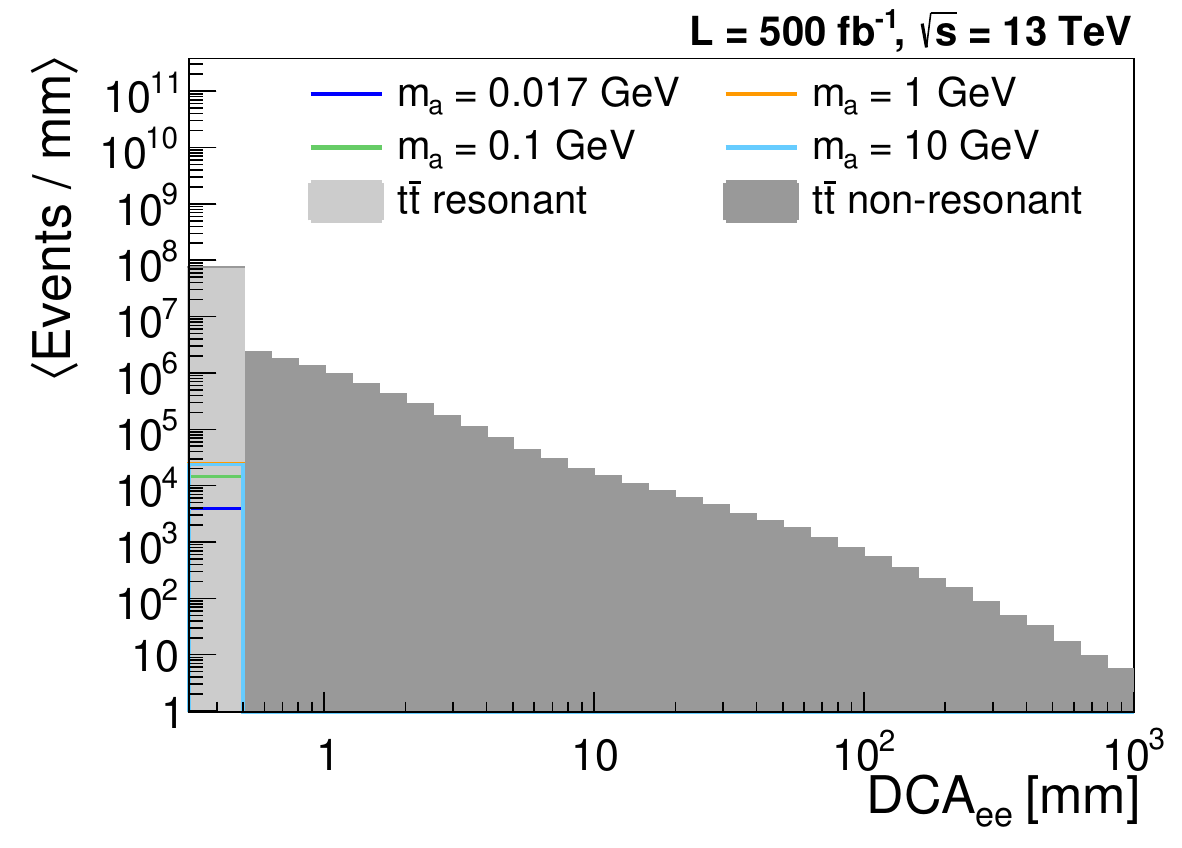}

    \vspace{0.5em}

    \includegraphics[width=0.48\linewidth,
        trim={15pt 0 5pt 0},
        clip]{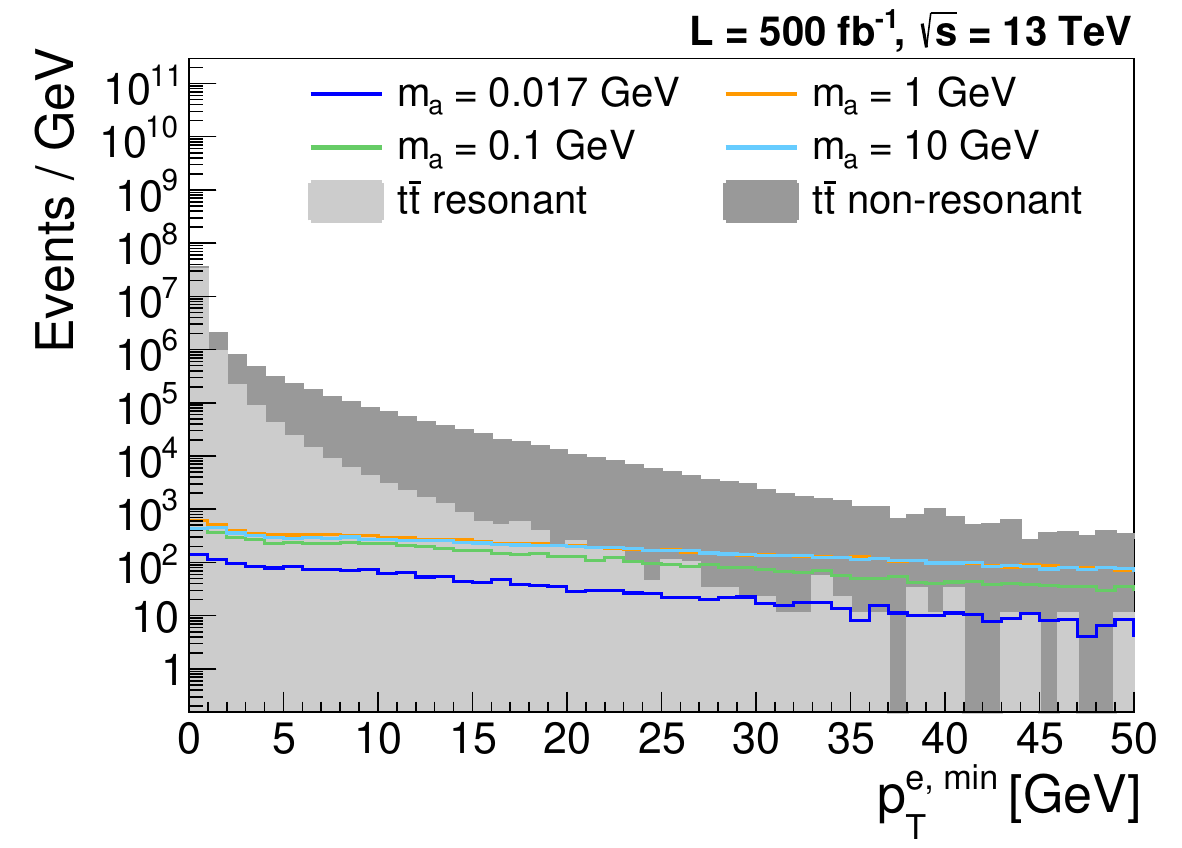}
    \hfill
    \includegraphics[width=0.48\linewidth,
        trim={15pt 0 5pt 0},
        clip]{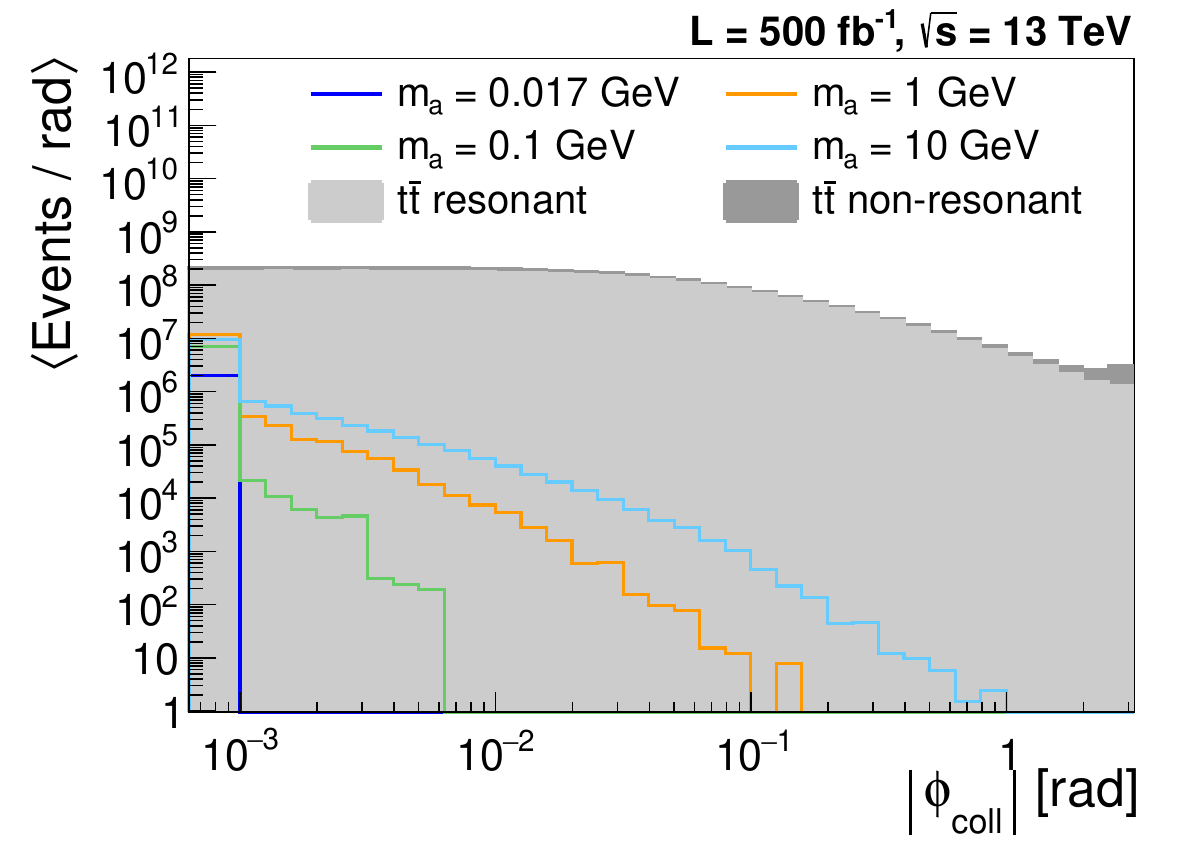}

    \caption{Distribution of the discriminating features used for further background suppression after applying the pre-selection and invariant-mass selection: the isolation of the least isolated electron in the pair (upper left), the distance of closest approach between the electron and the positron (upper right), the transverse momentum of the electron with the lowest $p_T$ in the pair (lower left), and the collinearity angle (lower right). For the distance of closest approach and the collinearity angle, the first bin contains all events below 0.5~mm and $10^{-3}$~rad, respectively, and the yields are divided by the bin width. The plots show several signal benchmarks with $c\tau = 1\,\,\mathrm{mm}$, together with the resonant (light gray) and non-resonant (dark gray) $t\bar{t}$ backgrounds. The background yield is normalized to its experimental cross section, while the signal yields are normalized to an illustrative cross section of $0.1\,\,\mathrm{pb}$. An integrated luminosity of $500\,\,\mathrm{fb^{-1}}$ is assumed.}
    \label{fig:variables}
\end{figure}

The discriminating variables used for further background suppression are summarized in Table~\ref{tab:variables}. Their distributions after the pre-selection and invariant-mass vetoes are shown in Fig.~\ref{fig:variables}.

The electron trajectories are constructed as straight lines from their generator-level production points along their momentum directions. The distance of closest approach is defined as the minimum three-dimensional separation between these trajectories. The secondary-vertex position entering the collinearity angle is defined as the point that minimizes the sum of the squared perpendicular distances to the two electron trajectories, treated as infinite straight lines. As discussed in Sec.~\ref{sec:strategy}, this treatment neglects the curvature of the electron trajectories in the CMS magnetic field.

\begin{table}[!htbp]
\begin{center}
    \begin{tabular}{|>{\centering\arraybackslash}m{3cm}|m{11cm}|}
        \hline
        \rowcolor{black!15} Feature & Description \\ \hline
        Electron isolation ($I_e^\mathrm{min}$) & $I_e=p_{\mathrm{T}}^e/\sum_{\Delta R<0.3} p_{\mathrm{T}}^{\mathrm{charged}}$, where the sum runs over all charged particles within $\Delta R<0.3$, including the electron itself but excluding the other electron in the pair. Suppresses electrons produced inside jets. \\ \hline
        Distance of closest approach ($\mathrm{DCA}_{ee}$) & Minimum 3D distance between the electron trajectories. Suppresses non-resonant background. \\ \hline
        Transverse momentum ($p_T^{e,\mathrm{min}}$) & Transverse momentum of the electron with the lower $p_T$ in the pair. Suppresses $\pi^0 \to e^+ e^- \gamma$ decays, where $\pi^0$ is produced predominantly in $K_S^0$ decays in the simulated $t\bar{t}$ background, with the remaining cases arising mainly from decays of strange baryons. \\ \hline
        Collinearity angle ($\phi_\mathrm{coll}$)& Angle between the dielectron transverse momentum vector and the transverse displacement vector, $\vec{l}_{xy} \equiv (x, y)_\mathrm{SV} - (x, y)_\mathrm{PV}$, where PV denotes the primary vertex (taken as the interaction point) and SV denotes the secondary vertex. Suppresses displaced pairs not pointing back to the PV. \\ \hline
    \end{tabular}
    \caption{Discriminating features used for further background suppression after the pre-selection and invariant-mass selection.}
    \label{tab:variables}

\end{center}
\end{table}

The choice of selection thresholds is guided by the Punzi figure of merit \cite{punzi2003sensitivity}
\begin{equation*}
    f\left(\epsilon_S, B\right) \equiv \dfrac{\epsilon_S}{\sqrt{B} + \dfrac{a}{2}},
\end{equation*}
with $a = 3$, where $B$ is the number of background events passing the multidimensional rectangular selection, normalized to the expected number of background events at an integrated luminosity of $500 \,\, \mathrm{fb}^{-1}$, and $\epsilon_S$ is the efficiency of the chosen cuts on signal. The cut parameter space defined by the electron isolation, the distance of closest approach between the two electrons, the transverse momentum of the lower-$p_T$ electron in the pair, and the collinearity angle is scanned separately for each ($m_a, c\tau$) benchmark. A single common selection is then chosen as an informed compromise among the benchmark-specific optima.

A collinearity angle of $|\phi_\mathrm{coll}| < 0.05 \,\, \mathrm{rad}$ and a maximum distance of closest approach between the two electrons of $0.5 \,\, \mathrm{mm}$ are required. For dielectron pairs with an invariant mass $m_{ee} \geq 1 \,\, \mathrm{GeV}$, each electron is required to have an isolation greater than $0.4$ and a transverse momentum of $p_T > 8 \,\, \mathrm{GeV}$. For $m_{ee} < 1 \,\,\mathrm{GeV}$, the isolation requirement is tightened to greater than $0.7$, while the transverse-momentum threshold is lowered to $1 \,\, \mathrm{GeV}$ in order to maximize sensitivity to lower ALP masses.

\begin{figure}[!htbp]
    \centering
    \includegraphics[width=0.61\linewidth, trim = 0 70pt 0 170pt, clip]{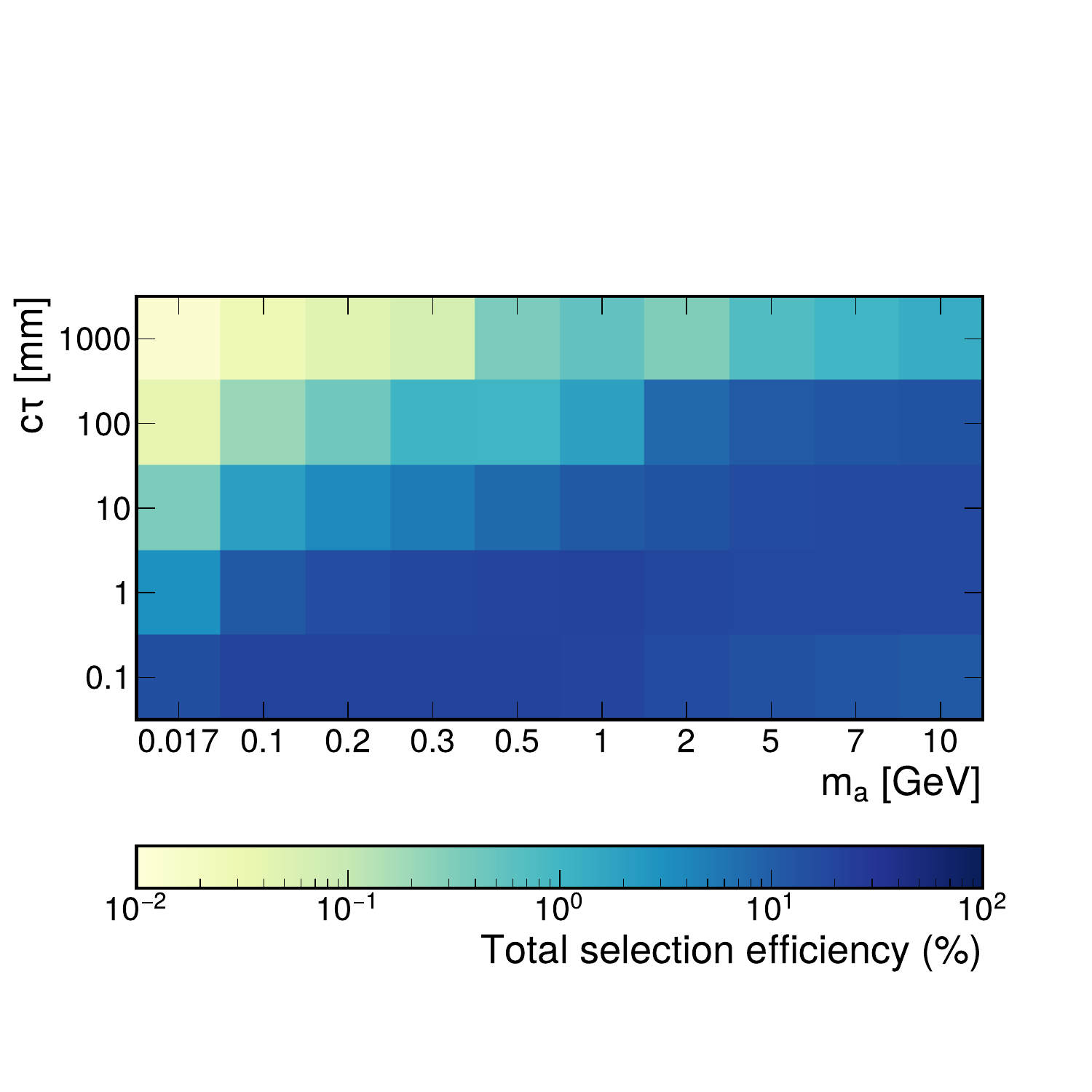}
    \caption{Total selection efficiencies for each signal benchmark, defined as the fraction of generated events that pass the pre-selection and full signal selection.}
    \label{fig:eff_sigselection}
\end{figure}

The full selection has an efficiency of $(1.84 \pm 0.02) \times 10^{-4}$ for the inclusive $t\bar{t}$ sample, corresponding to a suppression of approximately a factor of $500$ relative to the pre-selection. For the signal benchmarks with the highest projected sensitivity, the signal-selection stage retains more than 80\% of the events that pass the pre-selection. The total signal efficiencies, defined relative to the generated samples and including both the pre-selection and signal selection, are shown in Fig.~\ref{fig:eff_sigselection} and in Appendix~\ref{sec:selection_efficiencies}.

\section{Projected sensitivity}
\label{sec:sensitivity}

The sensitivity to the $t \bar{t}a$, $a \to e^+ e^-$ signature is evaluated by computing upper limits at $95\%$ confidence level (CL) on the product of the production cross section times the branching fraction of the ALP decay into dielectrons, $\sigma(pp \to t \bar{t} a)\times \mathcal{B}(a \to e^+ e^-)$, as a function of the ALP mass and proper decay length. As defined in Sec.~\ref{sec:simulation}, the quoted cross sections refer to the generated fiducial phase space with ALP $p_T > 20$~GeV and electron $|\eta|<3.5$.

To compute the expected upper limits, we use the \textsc{Combine} tool, a software package developed by the CMS Collaboration for statistical analyses~\cite{combine}. The expected limits are estimated using the frequentist \texttt{AsymptoticLimits} method~\cite{cowan2011asymptotic}.

\begin{figure}[!htbp]
    \centering
    \includegraphics[width=0.9\linewidth]{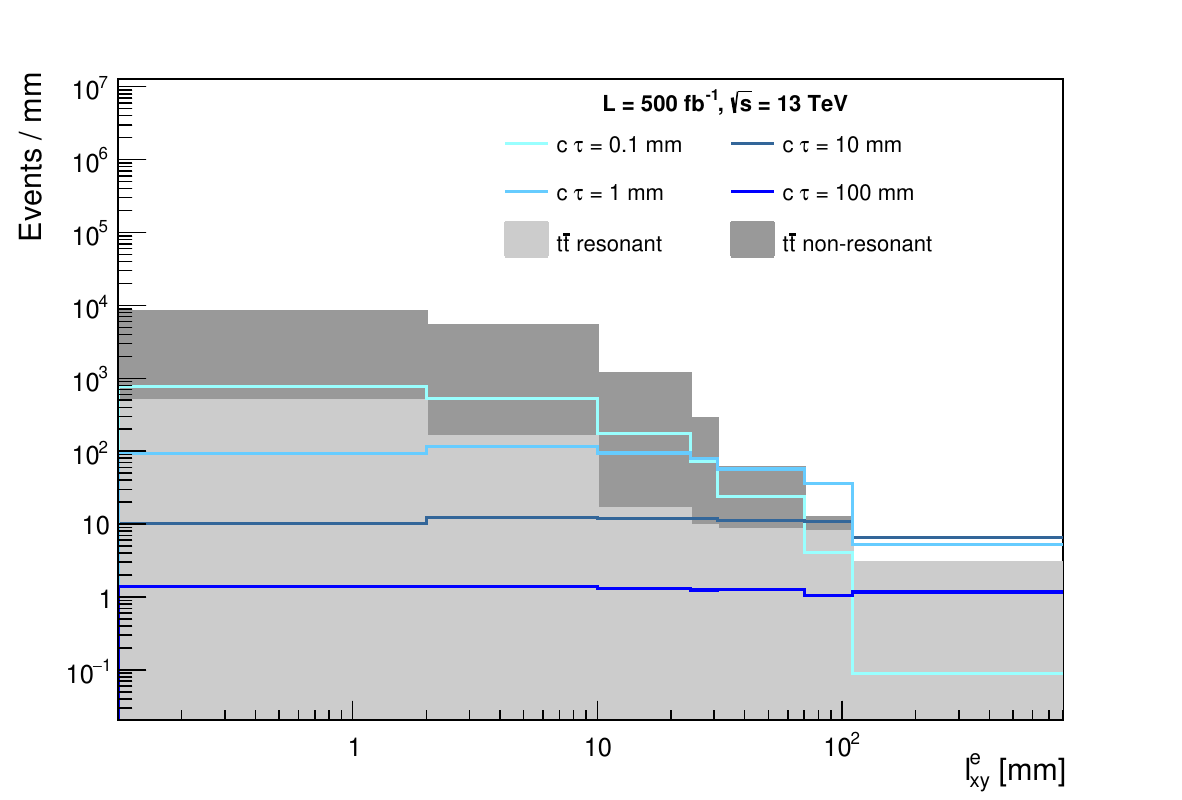}
    \caption{Distribution of the transverse displacement $l_{xy}$ of the least displaced electron after the full signal selection, showing several signal benchmarks with $m_a = 1 \,\, \mathrm{GeV}$, together with the resonant (light gray) and non-resonant (dark gray) $t\bar{t}$ backgrounds. The background yield is normalized to its experimental cross section, while the signal yields are normalized to an illustrative cross section of  $0.1 \,\, \mathrm{pb}$. An integrated luminosity of $500 \,\, \mathrm{fb^{-1}}$ is assumed.}
    \label{fig:displacement}
\end{figure}

To compute the expected upper limits, the events are categorized based on the displacement of the least displaced electron after the signal selection, using six bins corresponding to the pixel tracker and one bin for the outer tracker, as shown in Fig.~\ref{fig:displacement}. The displacement of the least displaced electron is used so that an event enters a higher-displacement category only when both electrons are at least that displaced, providing a conservative categorization for background pairs whose electrons can originate from different positions. For signal candidates, the two electrons originate from the same ALP decay point and therefore have the same generator-level $l_{xy}$. As a simplified treatment of the systematic uncertainties, independent 30\% rate uncertainties are applied to the signal and background yields. Each nuisance parameter is taken to be fully correlated across the seven $l_{xy}$ categories. 

\begin{figure}[htpb!]
    \centering
        \includegraphics[height=7.8cm, trim=1cm 0 0 0, clip]{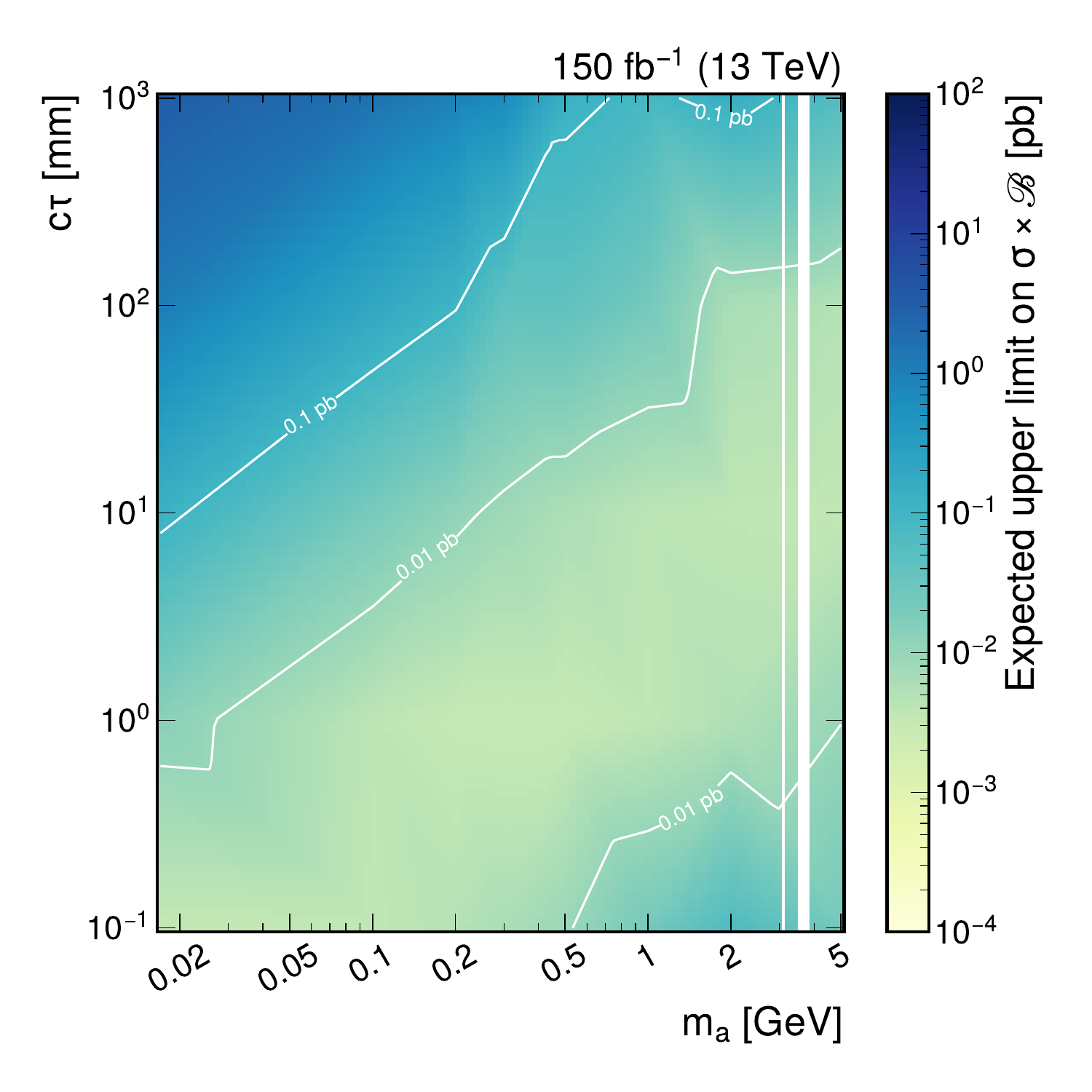}
        \includegraphics[height=7.8cm, trim=1cm 0 0 0, clip]{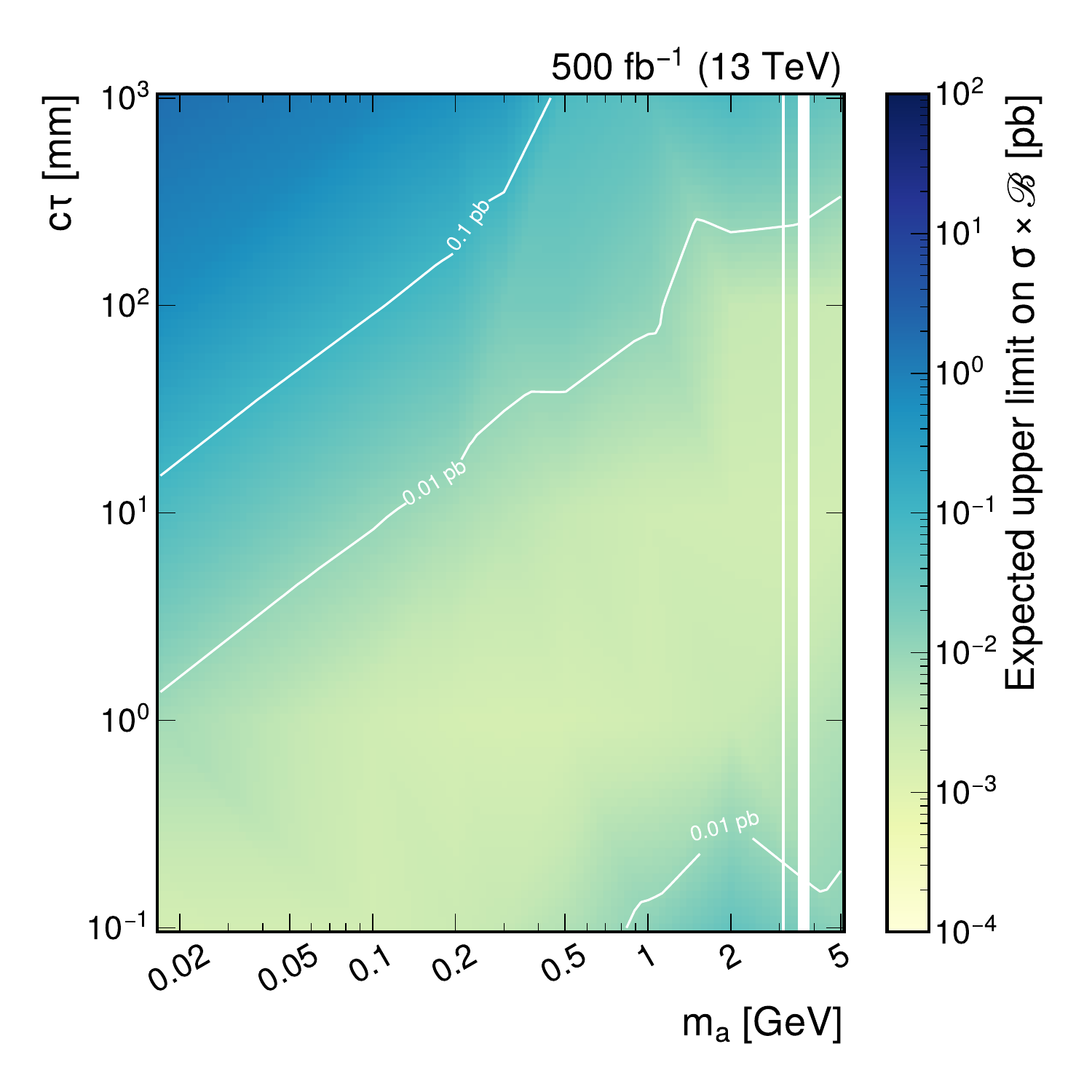}
    \caption{Expected 95\% CL upper limits on $\sigma(pp\to t\bar{t}a)\times\mathcal{B}(a\to e^+e^-)$ for integrated luminosities of $150 \,\, \mathrm{fb}^{-1}$ (left) and $500 \,\, \mathrm{fb}^{-1}$ (right) in the $(m_a, c\tau)$ parameter space, using a linear interpolation of the logarithm of the expected upper limit in $(\log m_a,\log c\tau)$ between the studied points. The white contours indicate selected values of the expected upper limit. The mass windows vetoed in the selection are excluded from the plotted range.}
    \label{fig:limits}
\end{figure}

The expected upper limits are computed for each signal benchmark. Between the simulated ($m_a, c\tau$) points, the limits are interpolated linearly in $\log m_a$, $\log c\tau$, and $\log[\sigma(pp\to t\bar t a) \times\mathcal{B}(a\to e^+e^-)]$.  Fig.~\ref{fig:limits} shows the expected limits for integrated luminosities of $150 \,\, \mathrm{fb}^{-1}$ and $500 \,\, \mathrm{fb}^{-1}$. These luminosity values are chosen to be representative of the integrated luminosity of the LHC Run~2 dataset and of the combined Run~2 and Run~3 datasets, respectively; both projections are evaluated at $\sqrt{s}=13$ TeV. The corresponding one-dimensional limits for each studied $c\tau$ value, including the expected $\pm1$ and $\pm2$ standard-deviation variations, are shown in Appendix~\ref{sec:1Dlimits}. At the shortest studied $c\tau$, the best limits are expected for the lower ALP masses, while for the largest proper decay lengths the limits degrade as the mass decreases. Both behaviors are mostly driven by the previously discussed boost dependence of the signal efficiency. In the most sensitive region of parameter space, values of $\sigma(pp \to t \bar{t} a)\times \mathcal{B}(a \to e^+ e^-)$ as small as $\mathcal{O}(10^{-3}) \,\, \mathrm{pb}$ can be probed. A model-specific interpretation of the limits in terms of ALP couplings is given in Appendix~\ref{sec:top-scenario}.

The projected sensitivity extends below the dimuon threshold and reaches the lowest masses considered in this study. For these low-mass benchmarks, however, the electron pairs are highly collimated, and the limits do not include a loss of efficiency associated with resolving and reconstructing two displaced electrons. The corresponding sensitivity should therefore be interpreted as an idealized generator-level projection.

This study assumes 100\% efficiency for the trigger, $t\bar{t}$ reconstruction, and displaced-dielectron reconstruction. To assess the impact of these assumptions, the combined efficiency for triggering on and reconstructing the prompt $t\bar{t}$ system used to select the event is reduced to 10\%, with the corresponding factor applied to both the signal and simulated $t\bar{t}$ background yields. The expected limits then weaken by approximately a factor of 3, close to the $S/\sqrt{B}$ scaling expected when signal and background are both reduced by a factor of 10 in a background-dominated regime. The sensitivity to a larger total background yield is probed separately by increasing the simulated background yields by factors of 10 and 100, with the signal unchanged; the expected limits weaken by approximately factors of 3 and 10, respectively. Finally, flat displaced-dielectron reconstruction efficiencies of 10\% and 1\% are applied to the signal only, with the prompt-event selection restored to its nominal efficiency and the background unchanged; the expected limits weaken by approximately factors of 10 and 100, respectively. For the variations considered, the projected sensitivity is therefore less strongly affected by the efficiency for selecting the event using the prompt $t\bar{t}$ system or by large changes in the background normalization than by the displaced-dielectron reconstruction efficiency, to which the limits scale approximately inversely. These variations are intended to test the robustness of the projection rather than to provide estimates of realistic detector-level efficiencies or background rates. The 30\% rate uncertainties described above are retained in all of these variations.

\section{Summary}
\label{sec:summary}
The sensitivity of a search for long-lived axion-like particles produced in association with top-quark pairs and decaying to displaced electron--positron pairs at the LHC has been investigated. The prompt objects from the $t\bar t$ system provide independent handles for triggering and event selection, allowing the event to be selected independently of the displaced electron pair. This strategy extends the top-associated LLP search program to masses below the dimuon threshold.

A generator-level search strategy motivated by the CMS detector was developed for ALP masses between 17~MeV and 10~GeV and proper decay lengths between 0.1~mm and 1~m. Inclusive SM $t\bar t$ simulation was used to model the physics background. Expected 95\% CL upper limits were computed for $\sigma(pp\to t\bar t a)\times\mathcal{B}(a\to e^+e^-)$ as a function of the ALP mass and proper decay length, for integrated luminosities of 150 and 500~fb$^{-1}$ at $\sqrt{s}=13$~TeV. In the most sensitive region of the parameter space considered, values of the cross section times branching fraction of order $10^{-3}$~pb can be probed for 500~fb$^{-1}$. The results should be interpreted as an idealized generator-level projection. Backgrounds arising from detector-material interactions, object misreconstruction, and other instrumental effects have not been simulated, and no detector-level model is included for the efficiency of reconstructing displaced electrons or resolving highly collimated electron pairs.

Robustness tests show that reducing the combined efficiency for triggering on and reconstructing the prompt $t\bar{t}$ system to 10\%, or increasing the simulated background yield by a factor of 10, weakens the expected limits by approximately a factor of 3; a hundredfold increase in the background weakens them by approximately a factor of 10. By contrast, flat displaced-dielectron reconstruction efficiencies of 10\% and 1\% weaken the limits by approximately factors of 10 and 100, respectively. Among the effects considered, the displaced-dielectron reconstruction efficiency therefore has the strongest impact on the projected cross section sensitivity. This is particularly relevant at the lowest ALP masses, where very small angular separations between the electron and positron are predicted and their trajectories can be significantly affected by their opposite bending in the CMS magnetic field. A detector-level study of these effects will therefore be required to establish the experimental sensitivity of a Run 2 and Run 3 analysis in this regime. For future HL-LHC searches, the fine granularity of the CMS high-granularity calorimeter~\cite{CMS:2017jpq}, to be installed for the Phase-2 upgrade, may provide additional information for resolving highly collimated electron pairs in the endcap region.

The results of this study quantify the generator-level potential of the top-associated strategy for displaced-dielectron signatures and provide a basis for a detector-level experimental study. This work establishes top-associated production as a promising strategy for probing light long-lived particles decaying to electron pairs, a regime largely unexplored at the LHC.


\acknowledgments
\enlargethispage{2\baselineskip}
JA, FB, CC, AP, LR, and RSR acknowledge support from DESY (Hamburg, Germany), a member of the Helmholtz Association HGF, and support by the Deutsche Forschungsgemeinschaft (DFG, German Research Foundation) under Germany’s Excellence Strategy -- EXC 2121 "Quantum Universe" -- 390833306. XCV and RSR acknowledge support from the Spanish Research State Agency under projects PID2022-139514NA-C33, PCI2023-145984-2 and EUR2025-164824; by the “María de Maeztu” grant CEX2023-001318-M, funded by MICIU/AEI/10.13039/501100011033; and by the Xunta de Galicia (CIGUS Network of Research Centres, project 2025-CP119 and project ED431B 2026/17).

\clearpage
\appendix

\section{Selection efficiencies} \label{sec:selection_efficiencies}
\medskip

\begingroup
\fontsize{9pt}{10pt}\selectfont
\renewcommand{\arraystretch}{0.99}

\centering
\begin{tabular}{|c|c|c|c|c|}
\hline
Process & $m_a \,\, \left[\mathrm{GeV} \right]$ & $c\tau \,\, [\mathrm{mm}]$ & Pre-selection efficiency (\%) & Total selection efficiency (\%) \\ \hline
$t\bar{t}$ & -- & -- & $9.019 \pm 0.005$ & $(1.84 \pm 0.02) \times 10^{-2}$ \\ \hline
\multirow{50}{*}{$t\bar{t}a$} & \multirow{5}{*}{$0.017$} & $10^{-1}$ & $18.144 \pm 0.086$ & $14.738 \pm 0.079$ \\ \cline{3-5} 
 & & $10^{0}$ & $3.997 \pm 0.044$ & $3.137 \pm 0.039$ \\ \cline{3-5} 
 & & $10^{1}$ & $0.434 \pm 0.015$ & $0.337 \pm 0.013$ \\ \cline{3-5} 
 & & $10^{2}$ & $0.046 \pm 0.005$ & $0.038 \pm 0.004$ \\ \cline{3-5} 
 & & $10^{3}$ & $0.016 \pm 0.002$ & $0.013 \pm 0.002$ \\ \cline{2-5}
 & \multirow{5}{*}{$0.1$} & $10^{-1}$ & $24.827 \pm 0.097$ & $20.642 \pm 0.091$ \\ \cline{3-5} 
 & & $10^{0}$ & $14.278 \pm 0.078$ & $11.517 \pm 0.071$ \\ \cline{3-5} 
 & & $10^{1}$ & $2.540 \pm 0.035$ & $2.037 \pm 0.032$ \\ \cline{3-5} 
 & & $10^{2}$ & $0.270 \pm 0.012$ & $0.210 \pm 0.010$ \\ \cline{3-5} 
 & & $10^{3}$ & $0.034 \pm 0.003$ & $0.027 \pm 0.003$ \\ \cline{2-5}
 & \multirow{5}{*}{$0.2$} & $10^{-1}$ & $25.027 \pm 0.097$ & $20.977 \pm 0.091$ \\ \cline{3-5} 
 & & $10^{0}$ & $19.456 \pm 0.089$ & $15.849 \pm 0.082$ \\ \cline{3-5} 
 & & $10^{1}$ & $4.698 \pm 0.047$ & $3.701 \pm 0.042$ \\ \cline{3-5} 
 & & $10^{2}$ & $0.541 \pm 0.016$ & $0.411 \pm 0.014$ \\ \cline{3-5} 
 & & $10^{3}$ & $0.058 \pm 0.004$ & $0.046 \pm 0.003$ \\ \cline{2-5}
 & \multirow{5}{*}{$0.3$} & $10^{-1}$ & $25.002 \pm 0.097$ & $20.966 \pm 0.091$ \\ \cline{3-5} 
 & & $10^{0}$ & $21.655 \pm 0.092$ & $17.835 \pm 0.086$ \\ \cline{3-5} 
 & & $10^{1}$ & $6.524 \pm 0.055$ & $5.208 \pm 0.050$ \\ \cline{3-5} 
 & & $10^{2}$ & $1.349 \pm 0.026$ & $1.053 \pm 0.023$ \\ \cline{3-5} 
 & & $10^{3}$ & $0.086 \pm 0.005$ & $0.067 \pm 0.004$ \\ \cline{2-5}
 & \multirow{5}{*}{$0.5$} & $10^{-1}$ & $24.682 \pm 0.096$ & $20.759 \pm 0.091$ \\ \cline{3-5} 
 & & $10^{0}$ & $23.614 \pm 0.095$ & $19.665 \pm 0.089$ \\ \cline{3-5} 
 & & $10^{1}$ & $9.485 \pm 0.066$ & $7.605 \pm 0.059$ \\ \cline{3-5} 
 & & $10^{2}$ & $1.312 \pm 0.025$ & $1.027 \pm 0.023$ \\ \cline{3-5} 
 & & $10^{3}$ & $0.438 \pm 0.010$ & $0.345 \pm 0.009$ \\ \cline{2-5}
 & \multirow{5}{*}{$1$} & $10^{-1}$ & $23.817 \pm 0.095$ & $19.455 \pm 0.089$ \\ \cline{3-5} 
 & & $10^{0}$ & $24.840 \pm 0.097$ & $20.146 \pm 0.090$ \\ \cline{3-5} 
 & & $10^{1}$ & $14.459 \pm 0.079$ & $11.164 \pm 0.070$ \\ \cline{3-5} 
 & & $10^{2}$ & $2.487 \pm 0.035$ & $1.845 \pm 0.030$ \\ \cline{3-5} 
 & & $10^{3}$ & $0.718 \pm 0.013$ & $0.525 \pm 0.011$ \\ \cline{2-5}
 & \multirow{5}{*}{$2$} & $10^{-1}$ & $22.479 \pm 0.093$ & $16.471 \pm 0.083$ \\ \cline{3-5} 
 & & $10^{0}$ & $25.120 \pm 0.097$ & $18.067 \pm 0.086$ \\ \cline{3-5} 
 & & $10^{1}$ & $19.354 \pm 0.088$ & $13.017 \pm 0.075$ \\ \cline{3-5} 
 & & $10^{2}$ & $11.897 \pm 0.072$ & $7.896 \pm 0.060$ \\ \cline{3-5} 
 & & $10^{3}$ & $0.539 \pm 0.012$ & $0.311 \pm 0.009$ \\ \cline{2-5}
 & \multirow{5}{*}{$5$} & $10^{-1}$ & $18.768 \pm 0.087$ & $13.961 \pm 0.077$ \\ \cline{3-5} 
 & & $10^{0}$ & $24.268 \pm 0.096$ & $17.253 \pm 0.084$ \\ \cline{3-5} 
 & & $10^{1}$ & $23.628 \pm 0.095$ & $16.340 \pm 0.083$ \\ \cline{3-5} 
 & & $10^{2}$ & $16.317 \pm 0.083$ & $10.871 \pm 0.070$ \\ \cline{3-5} 
 & & $10^{3}$ & $1.253 \pm 0.018$ & $0.731 \pm 0.013$ \\ \cline{2-5}
 & \multirow{5}{*}{$7$} & $10^{-1}$ & $16.683 \pm 0.083$ & $12.747 \pm 0.075$ \\ \cline{3-5} 
 & & $10^{0}$ & $24.194 \pm 0.096$ & $17.405 \pm 0.085$ \\ \cline{3-5} 
 & & $10^{1}$ & $23.341 \pm 0.095$ & $16.657 \pm 0.083$ \\ \cline{3-5} 
 & & $10^{2}$ & $17.886 \pm 0.086$ & $12.277 \pm 0.073$ \\ \cline{3-5} 
 & & $10^{3}$ & $1.740 \pm 0.021$ & $1.019 \pm 0.016$ \\ \cline{2-5}
 & \multirow{5}{*}{$10$} & $10^{-1}$ & $14.419 \pm 0.079$ & $11.175 \pm 0.070$ \\ \cline{3-5} 
 & & $10^{0}$ & $23.424 \pm 0.095$ & $17.146 \pm 0.084$ \\ \cline{3-5} 
 & & $10^{1}$ & $23.456 \pm 0.095$ & $17.053 \pm 0.084$ \\ \cline{3-5} 
 & & $10^{2}$ & $19.132 \pm 0.088$ & $13.325 \pm 0.076$ \\ \cline{3-5} 
 & & $10^{3}$ & $2.206 \pm 0.023$ & $1.344 \pm 0.018$ \\ \hline
\end{tabular}
\captionof{table}{Pre-selection and total selection efficiencies. The uncertainties are statistical.}
\label{tab:efficiencies}
\endgroup

\clearpage

\section{Expected upper limits for each studied \texorpdfstring{$c\tau$}{ctau} value} \label{sec:1Dlimits}
    \centering
    \begin{minipage}{0.47\linewidth}
        \includegraphics[width=0.91\textwidth, trim = 0 0 0 85pt, clip]{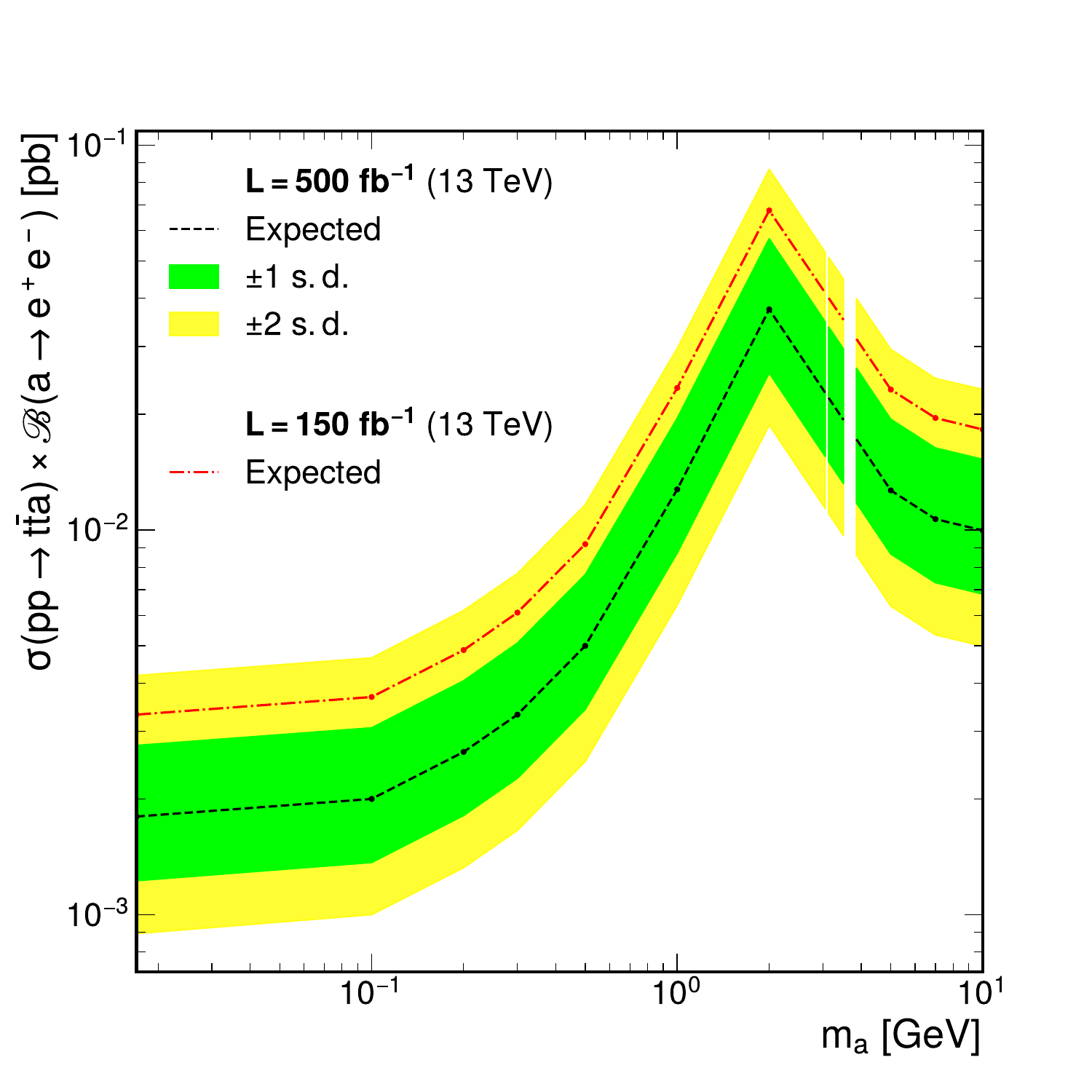}
    \vspace{-9pt}\\
    \centering{(a) $c\tau = 0.1 \,\, \mathrm{mm}$}
    \end{minipage}
    \begin{minipage}{0.47\linewidth}
        \includegraphics[width=0.91\textwidth, trim = 0 0 0 85pt, clip]{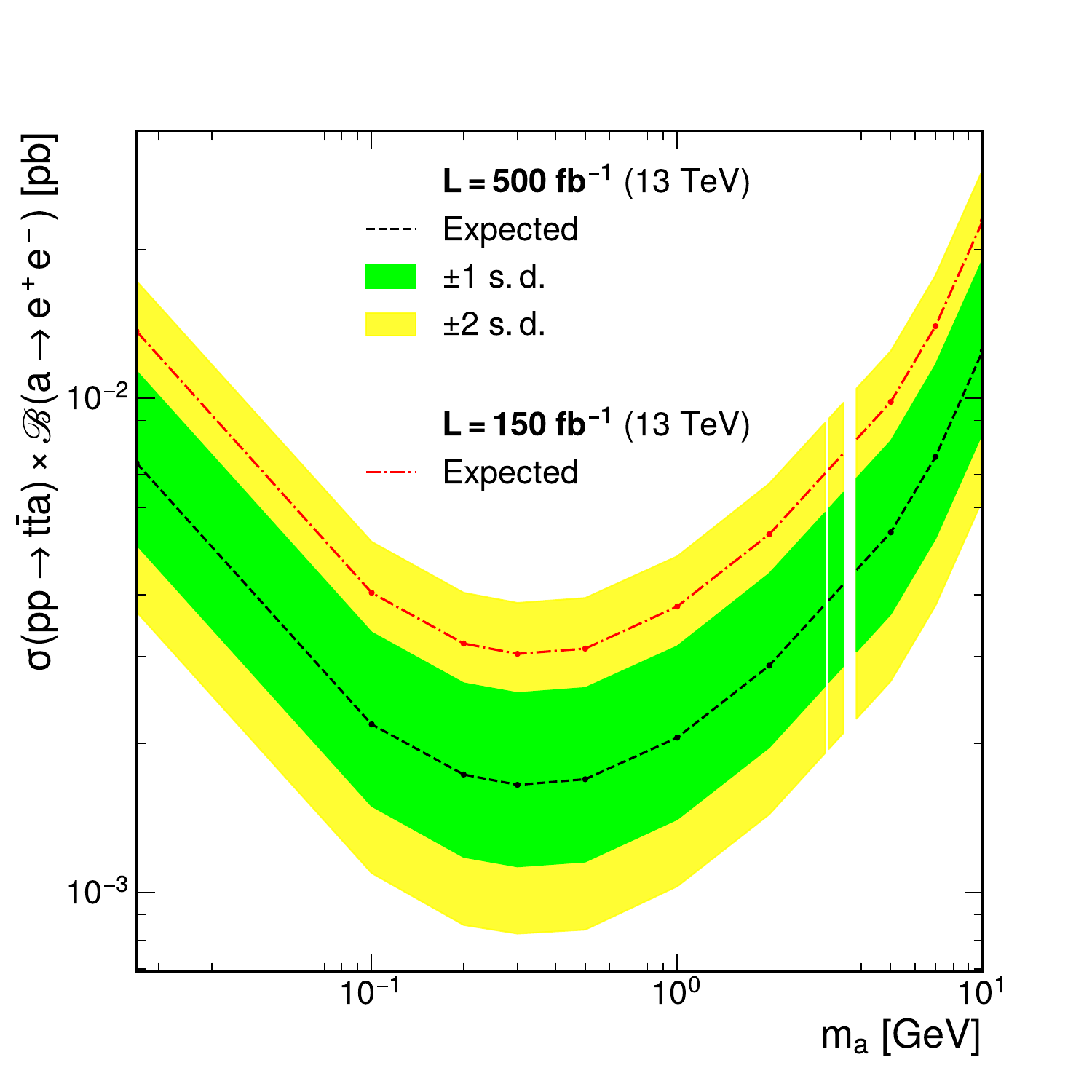}
    \vspace{-9pt}\\
    \centering{(b) $c\tau = 1 \,\, \mathrm{mm}$}
    \end{minipage}
    \vspace{10pt}\\
    
    \includegraphics[width=0.4277\textwidth, trim = 0 0 0 85pt, clip]{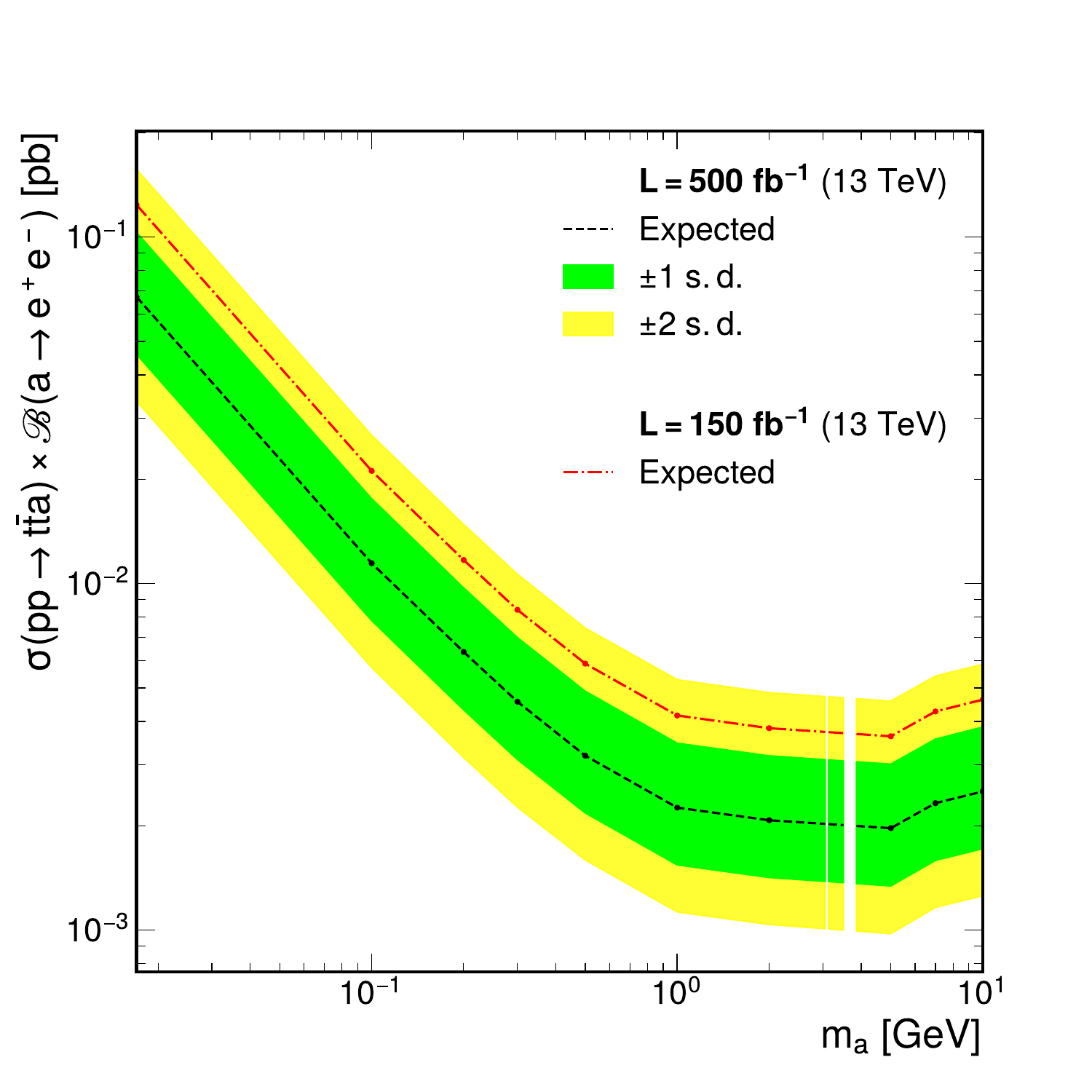}
    \vspace{-9pt}\\
    \centering{(c) $c\tau = 10 \,\, \mathrm{mm}$}
    \vspace{10pt}\\

    \begin{minipage}{0.47\linewidth}
    \includegraphics[width=0.91\textwidth, trim = 0 0 0 85pt, clip]{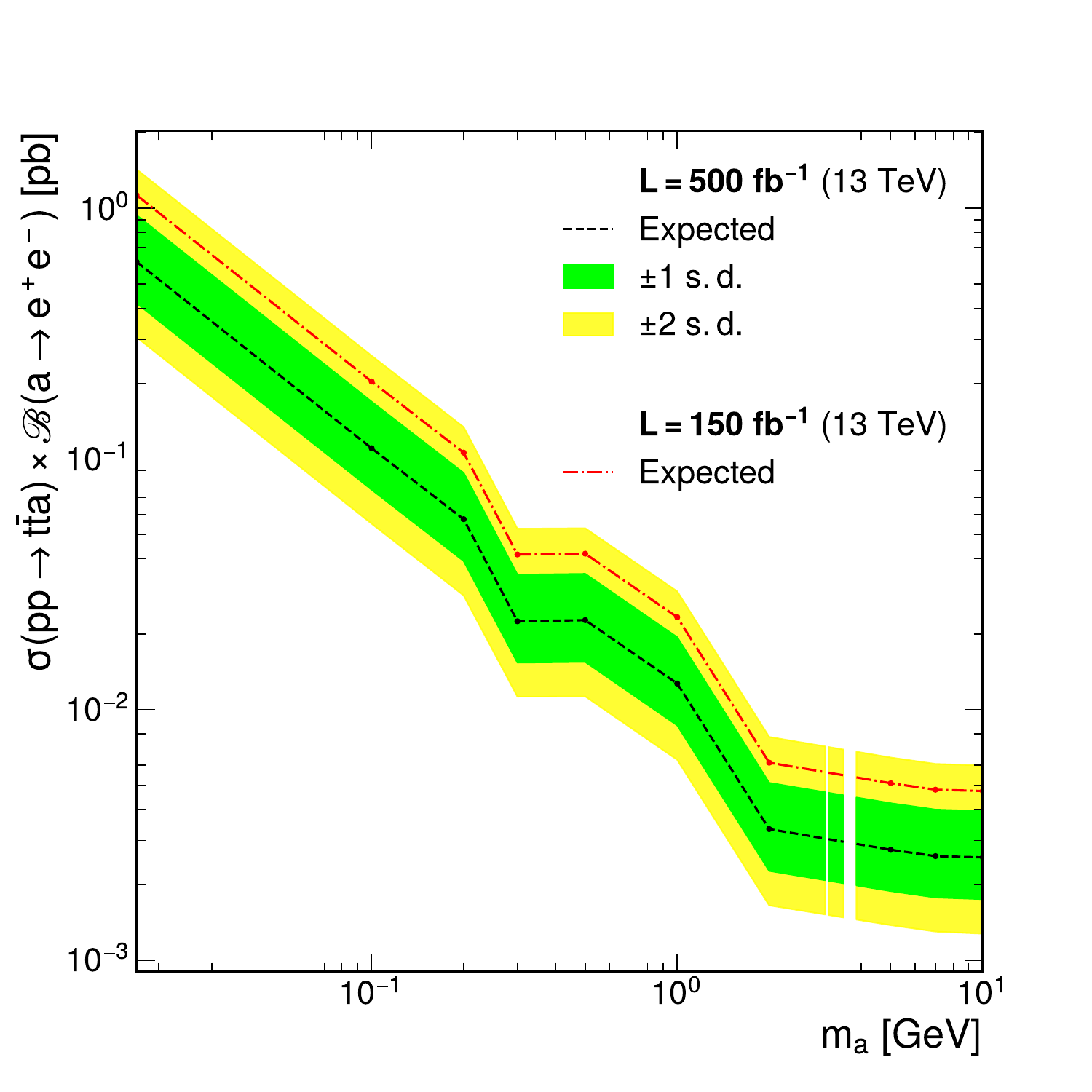}
    \vspace{-9pt}\\
    \centering{(d) $c\tau = 100 \,\, \mathrm{mm}$}
    \end{minipage}
    \begin{minipage}{0.47\linewidth}
    \includegraphics[width=0.91\textwidth, trim = 0 0 0 85pt, clip]{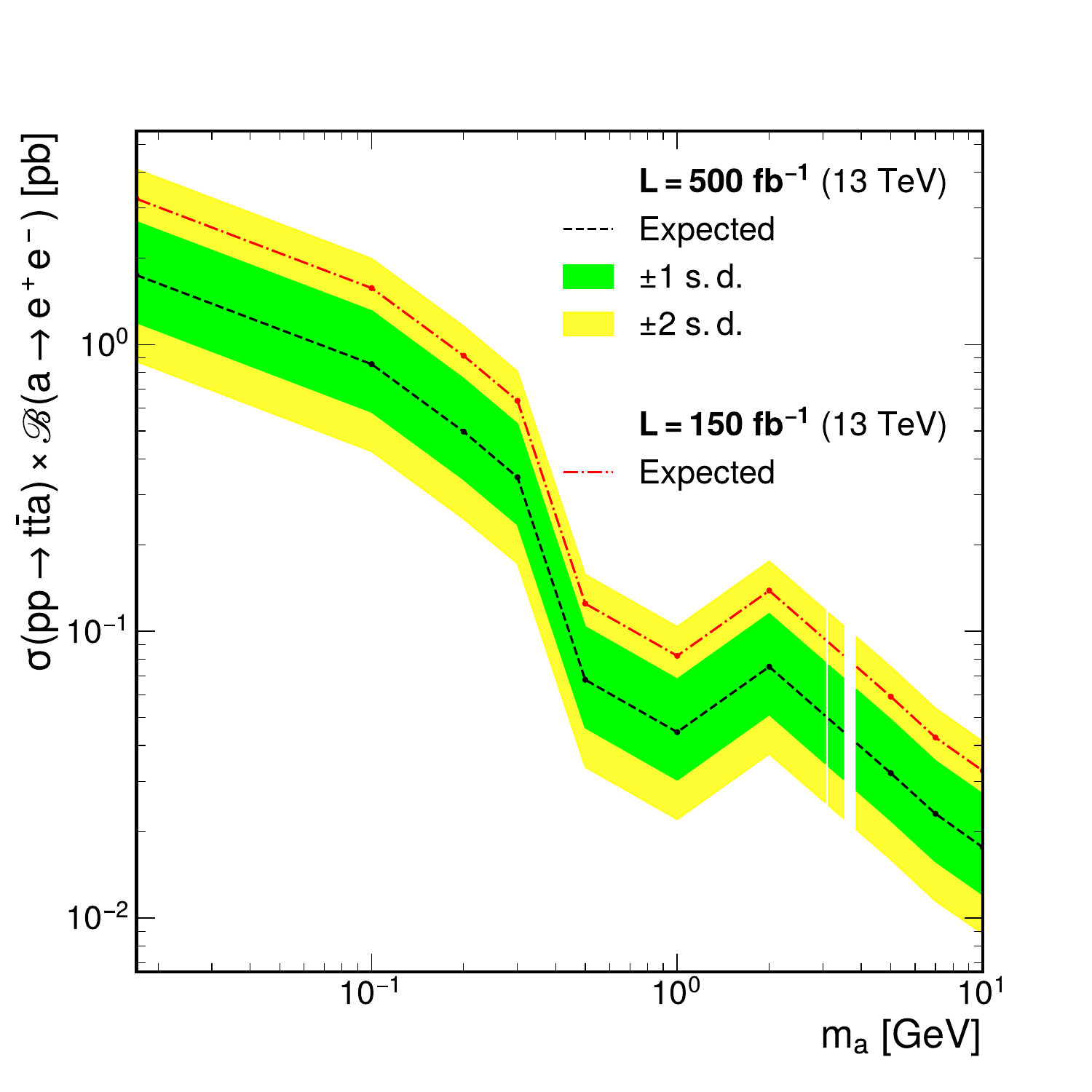}
    \vspace{-9pt}\\
    \centering{(e) $c\tau = 1000 \,\, \mathrm{mm}$}
    \end{minipage}
    \captionof{figure}{Expected $95\%$ CL upper limits on $\sigma(pp \to t \bar{t} a) \times \mathscr{B}(a \to e^+ e^-)$ as a function of the ALP mass $m_a$, for the studied $c\tau$ values and integrated luminosities of 150 and 500~fb$^{-1}$ at $\sqrt{s}=13$~TeV. The bands show the expected variations within one and two standard deviations (s.d.) for the 500~fb$^{-1}$ projection. The mass windows vetoed in the selection are excluded from the plotted range.}
    \label{fig:limits_by_ctau}
\justifying

\section{Interpretation in the top scenario}
\label{sec:top-scenario}

Although the primary results of this study are presented as limits on $\sigma(pp\to t\bar{t}a)\times\mathcal{B}(a\to e^+e^-)$ as a function of the ALP mass $m_a$ and proper decay length $c\tau$, they can also be interpreted in a specific ALP benchmark model. We consider the \textit{top scenario} introduced in Ref.~\cite{rygaard2023}, in which the ALP coupling to top quarks is the only nonzero ALP coupling at the cutoff scale $\Lambda = 4\pi f_a$. In the notation of Ref.~\cite{rygaard2023}, this corresponds to
\begin{equation}
    c_{tt}(\Lambda) \equiv c_{tt},
    \qquad
    c_{ff}(\Lambda) = 0 \quad (f \neq t),
    \qquad
    c_{VV}(\Lambda) = 0,
\end{equation}
where $c_{ff}$ and $c_{VV}$ denote the ALP couplings to SM fermions and gauge fields, respectively. We set $f_a = 1 \,\, \mathrm{TeV}$, following Ref.~\cite{rygaard2023}. The phenomenology is then specified by two parameters: the ALP mass $m_a$ and the effective top coupling $|c_{tt}|/f_a$.

Renormalization-group evolution and loop effects generate ALP couplings to lighter fermions and gauge bosons at lower energy scales. Numerical predictions for the induced couplings and decay widths are obtained using the public \texttt{ttalps} code~\cite{ttalps}, which implements the results of Ref.~\cite{bauer2021low,Bauer:2021wjo}. In the top scenario, and for fixed $f_a$, the induced couplings scale linearly with $|c_{tt}|/f_a$. At fixed $m_a$ and $f_a$, the partial and total decay widths therefore scale as $(c_{tt}/f_a)^2$, while the proper decay length scales as $(f_a/c_{tt})^2$. The branching fractions are independent of the overall normalization of $c_{tt}$ for fixed $f_a$. Additional ALP couplings at the cutoff scale would in general modify the decay width and branching fractions, so the relation between production rate, lifetime, and decay modes used here is specific to the top scenario.

\begin{figure}[!htpb]
    \centering
    \begin{subfigure}{0.47\linewidth}
        \centering
        \includegraphics[width=\linewidth]{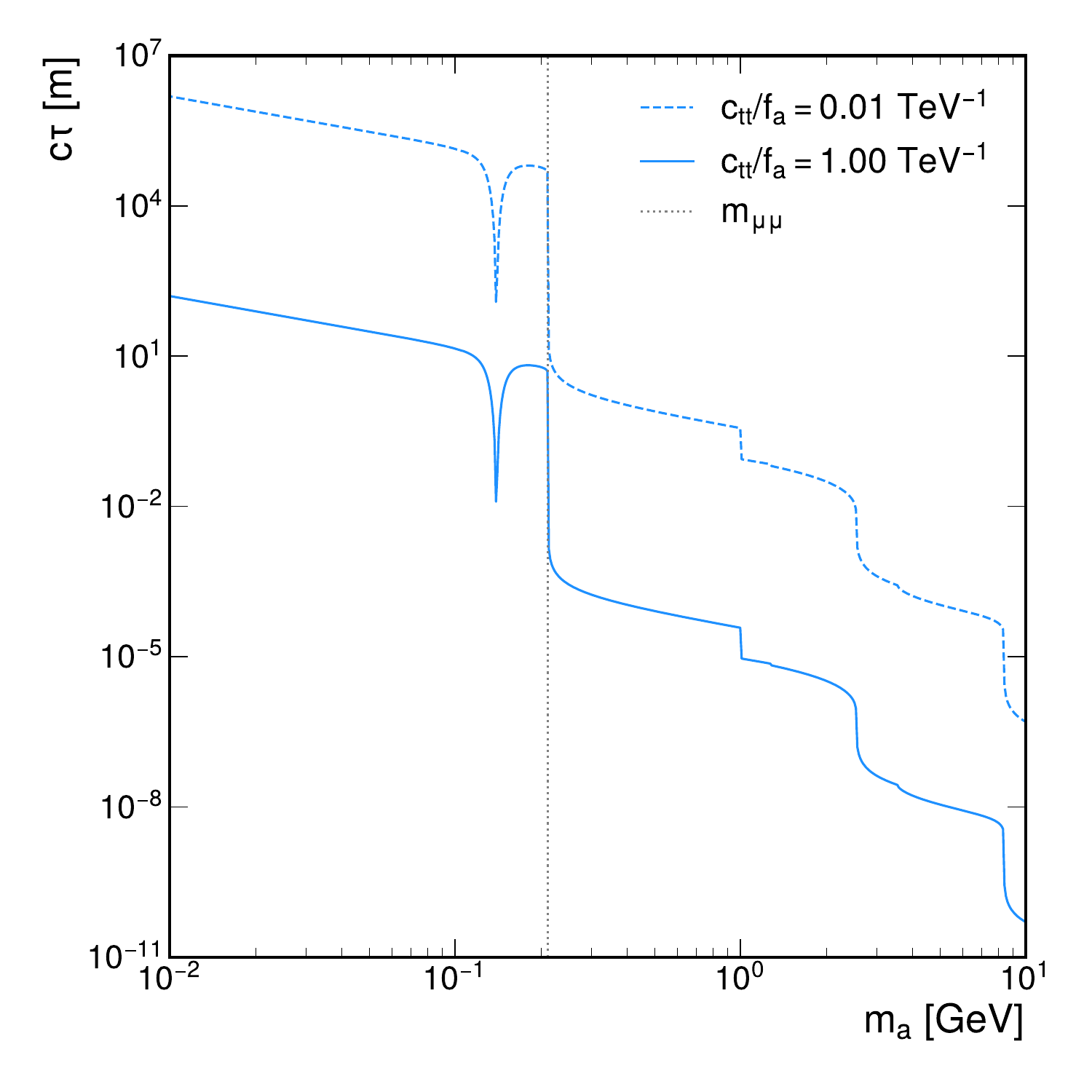}
    \end{subfigure}
    \hfill
    \begin{subfigure}{0.47\linewidth}
        \centering
        \includegraphics[width=\linewidth]{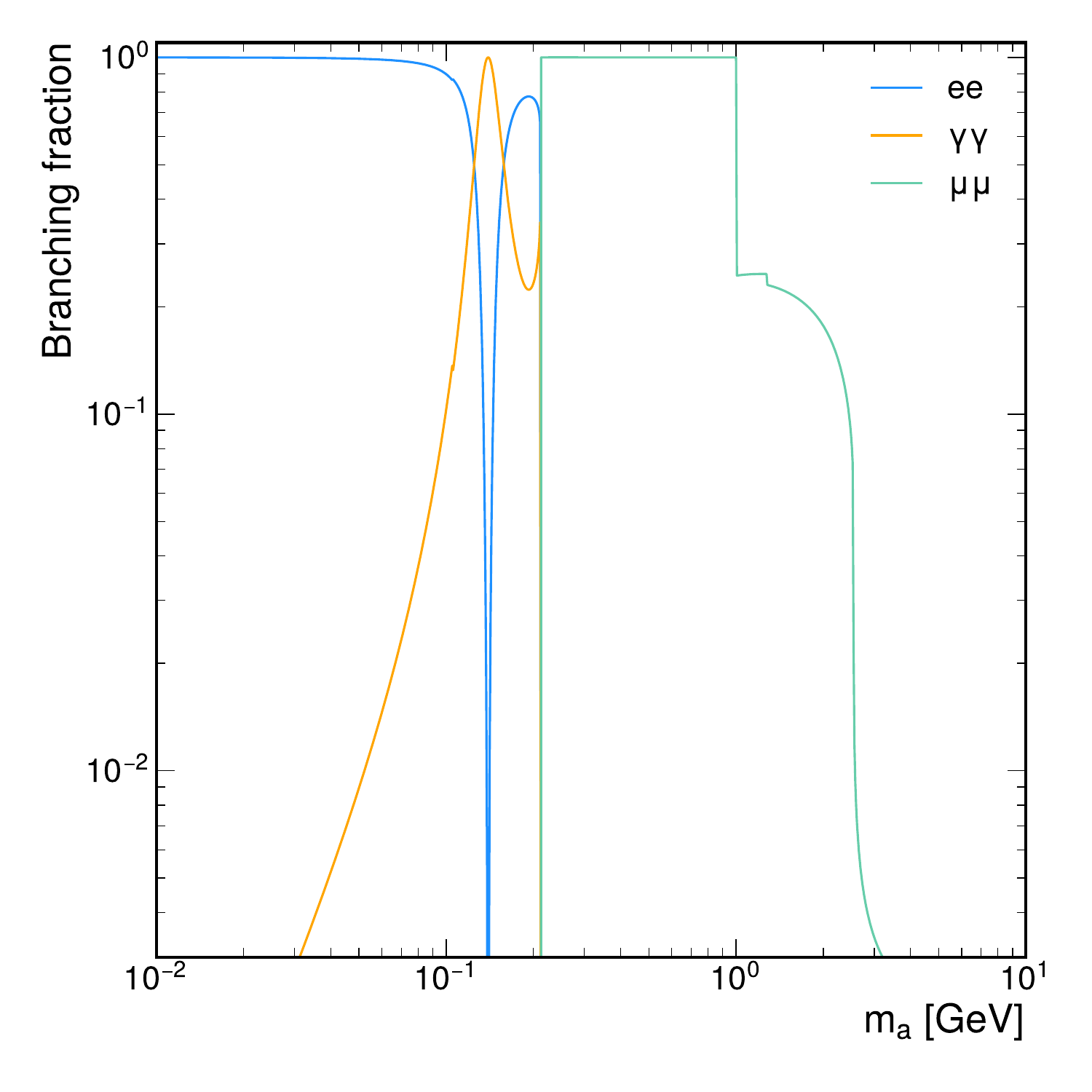}
    \end{subfigure}
    
    \caption{ALP proper decay length (left) and branching fractions to dielectrons, diphotons, and dimuons (right), calculated in the top scenario, as functions of the ALP mass. The dip in $c\tau$ around $m_a = m_\pi$ is due to ALP–pion mixing. The branching fraction to dielectrons is strongly suppressed above the dimuon threshold. For masses $m_a \geq 3 m_\pi$, hadronic channels open, and the branching fraction to dimuons is suppressed.}
    \label{fig:ctau_and_br}
\end{figure}

Figure~\ref{fig:ctau_and_br} shows the resulting ALP proper decay length as a function of the ALP mass for two representative $c_{tt} / f_a$ values (left), together with the branching fractions to dielectrons, diphotons, and dimuons (right). Below the dimuon threshold, only the dielectron and diphoton decay modes are kinematically accessible. At larger masses, the opening of additional leptonic and hadronic decay channels reduces $\mathcal{B}(a\to e^+e^-)$. Around the GeV scale, the branching-fraction predictions depend on the treatment of hadronic ALP decays; we follow the implementation used in Ref.~\cite{rygaard2023}.

The $t\bar{t}a$ production cross section is calculated with \texttt{MadGraph5\_aMC@NLO} for several ALP masses and coupling values. The cross sections are evaluated within the same generated fiducial phase space used for the primary results, requiring $p_T(a)>20~\mathrm{GeV}$ and $|\eta(e)|<3.5$. For fixed $f_a$, the cross section scales as $c_{tt}^2$, and its residual mass dependence is parameterized for the mass range considered using the generated samples as
\begin{equation}
    \sigma(pp\to t\bar{t}a)
    =
    (p_0 + p_1 m_a)\,c_{tt}^2. \label{eq:fit}
\end{equation}

\begin{figure}[h]
    \centering
    \includegraphics[width=0.55\linewidth]{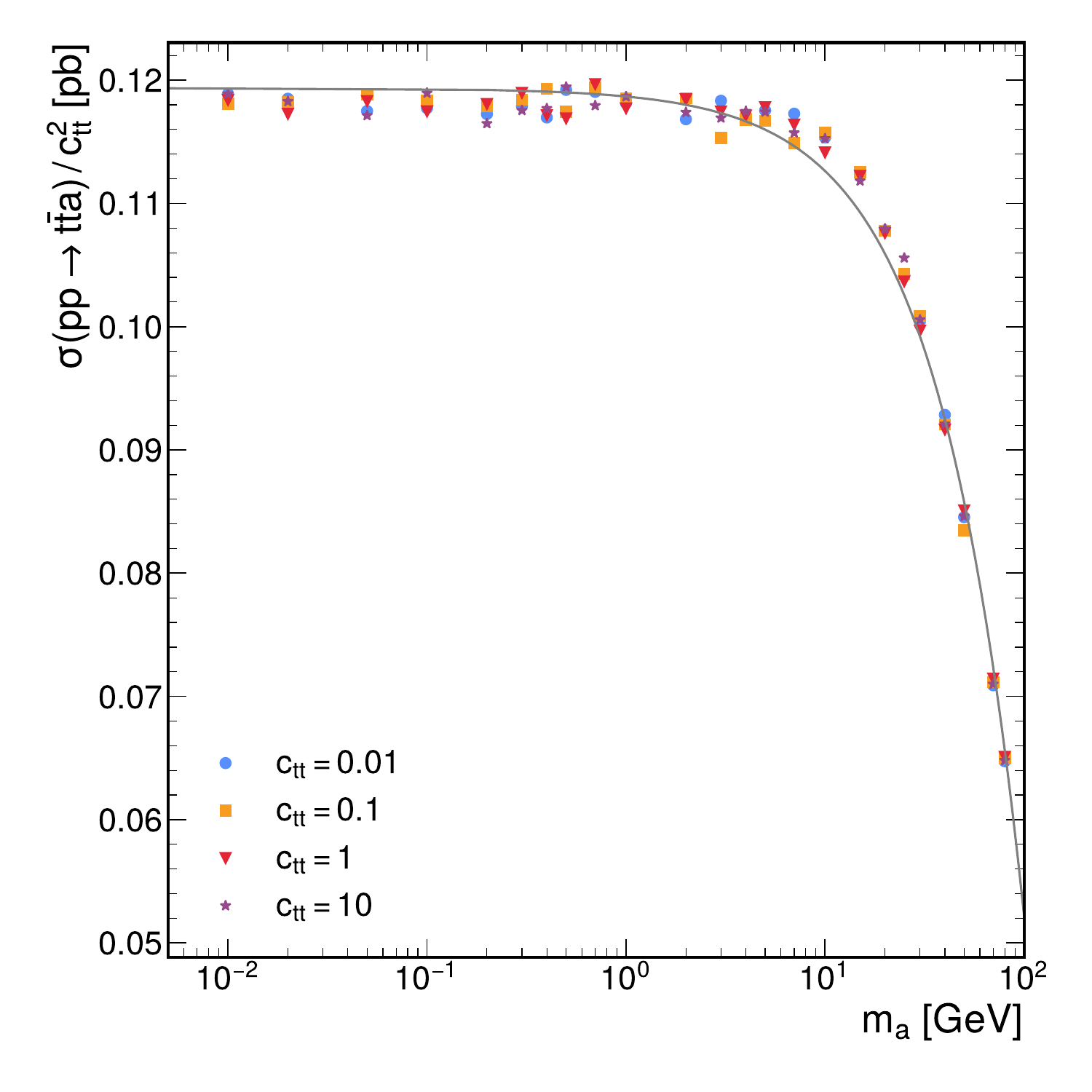}
    \caption{Dependence of $\sigma(pp\to t\bar t a)/c_{tt}^2$ on the ALP mass for the generated coupling values. The line shows the linear fit used in Eq.~\eqref{eq:fit}.}
    \label{fig:xsection_vs_coupling}
\end{figure}

The dependence of $\sigma(pp\to t\bar t a)/c_{tt}^2$ on the ALP mass is shown in Fig.~\ref{fig:xsection_vs_coupling} for the generated coupling values, together with the linear fit used in Eq.~\eqref{eq:fit}. The cross section is evaluated at the following mass points:
\begin{equation}
\begin{split}
m_a \in \,\,\{ & 0.01,\ 0.1,\ 0.2,\ 0.3,\ 0.4,\ 0.5,\ 0.7,\ 1,\ 2,\\
&3,\ 4,\ 5,\ 7,\ 10,\ 20,\ 40,\ 50,\ 70,\ 80\}~\mathrm{GeV}.
\end{split}
\end{equation}The fit gives $p_0 = (0.11935 \pm 0.00039) \,\, \mathrm{pb}$, $p_1 = (-6.71 \pm 0.14)\times 10^{-4} \,\, \mathrm{pb \,\, GeV}^{-1}$, where $m_a$ is expressed in GeV. The maximum fractional deviation between the fitted parameterization and the generated cross sections is approximately 3\%. No additional theoretical uncertainty from renormalization- and factorization-scale or PDF variations is included in the production cross section prediction.

Together with the predicted $\mathcal{B}(a\to e^+e^-)$ and $c\tau$, this provides a mapping from each point in the $(m_a,|c_{tt}|/f_a)$ parameter space to the corresponding $(m_a,c\tau,\sigma\mathcal{B})$ point constrained by the primary analysis, where $\sigma\mathcal{B}\equiv \sigma(pp\to t\bar t a) \times \mathcal{B}(a\to e^+e^-)$.

\begin{figure}[h]
    \centering
    \includegraphics[height=8.3cm, trim=1cm 0 0 0, clip]{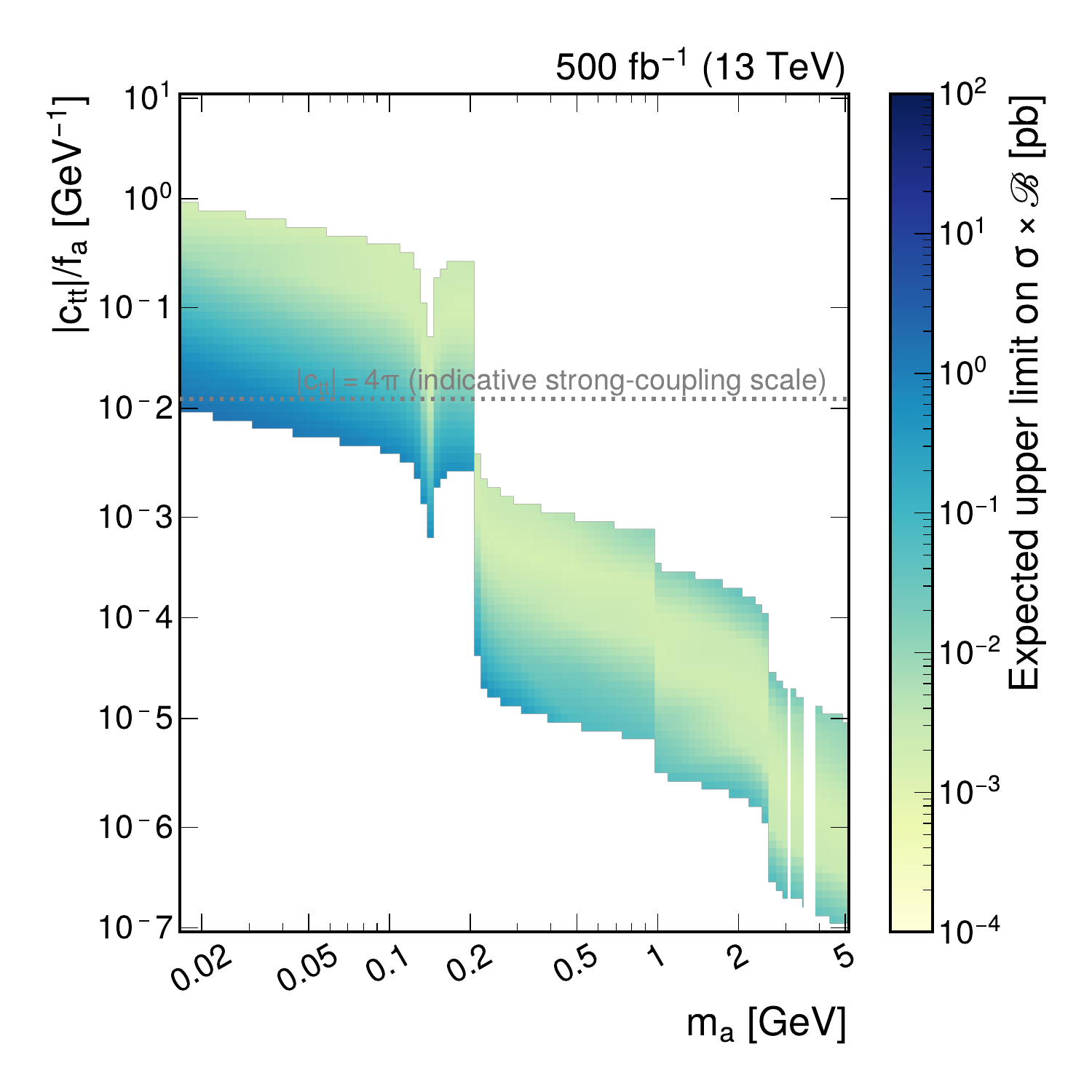}
    \includegraphics[height=8.3cm, trim=0 0 4.7cm 0, clip]{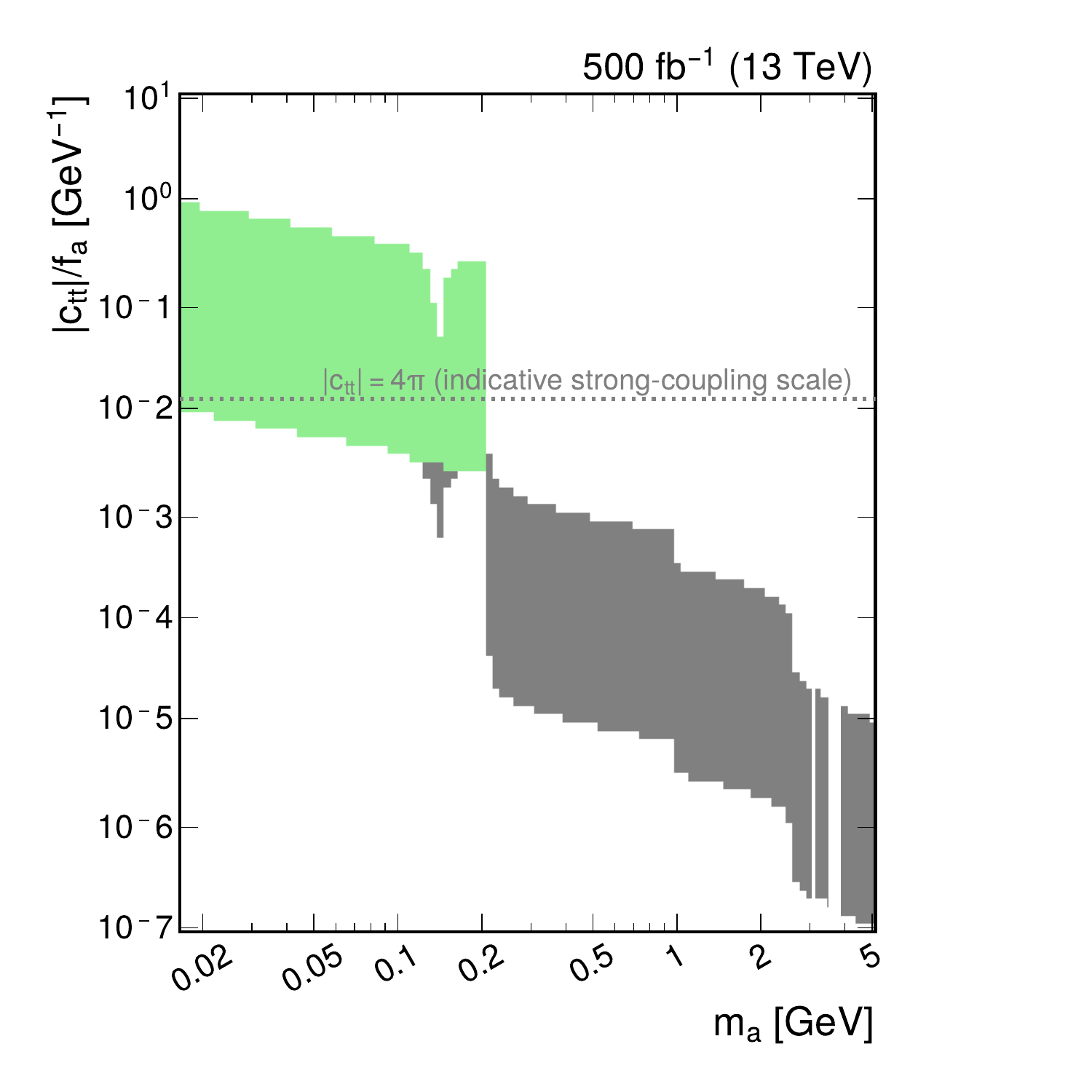}
    \caption{Projected sensitivity in the top scenario, evaluated for an integrated luminosity of $500$~fb$^{-1}$ at $\sqrt{s}=13$ TeV. Expected 95\% CL upper limits on $\sigma(pp\to t\bar{t}a)\times\mathcal{B}(a\to e^+e^-)$ (left) and exclusion region (right) in the $(m_a,|c_{tt}|/f_a)$ parameter space ($f_a = 1 \,\, \mathrm{TeV}$). The green region indicates the values of $m_a$ and $|c_{tt}|/f_a$ for which the theoretical prediction is above the expected upper limit. The gray region indicates the evaluated parameter space that is not excluded. The mass windows vetoed in the selection are excluded from the plotted range.}
    \label{fig:topscenario_exclusion}
\end{figure}

To obtain the projected sensitivity in the top scenario, the expected upper limits on $\sigma(pp\to t\bar{t}a)\times\mathcal{B}(a\to e^+e^-)$ are interpolated over the $(m_a,c\tau)$ plane using the logarithmic
interpolation described in Sec.~\ref{sec:sensitivity}. No extrapolation is performed outside the studied $(m_a,c\tau)$ range. For each point in the $(m_a,|c_{tt}|/f_a)$ parameter space, the predicted proper decay length determines the corresponding interpolated upper limit. The resulting expected upper limit on $\sigma(pp\to t\bar{t}a)\times\mathcal{B}(a\to e^+e^-)$ is shown in Fig.~\ref{fig:topscenario_exclusion} (left). Comparing this limit with the corresponding theoretical prediction gives the projected exclusion shown in Fig.~\ref{fig:topscenario_exclusion} (right). For the fixed choice $f_a=1$ TeV, part of the parameter space shown in Fig.~\ref{fig:topscenario_exclusion} corresponds to large values of the dimensionless coefficient $c_{tt}$. In this strong-coupling regime, the perturbative interpretation of the effective field theory and the quadratic extrapolation of the production cross section should be treated with caution. The precise boundary of this regime is model dependent; the large-$c_{tt}$ region is therefore shown to illustrate the mapping of the generator-level sensitivity rather than as a quantitatively robust effective field theory exclusion.

\begin{figure}[h]
    \centering
    \includegraphics[width=0.55\linewidth]{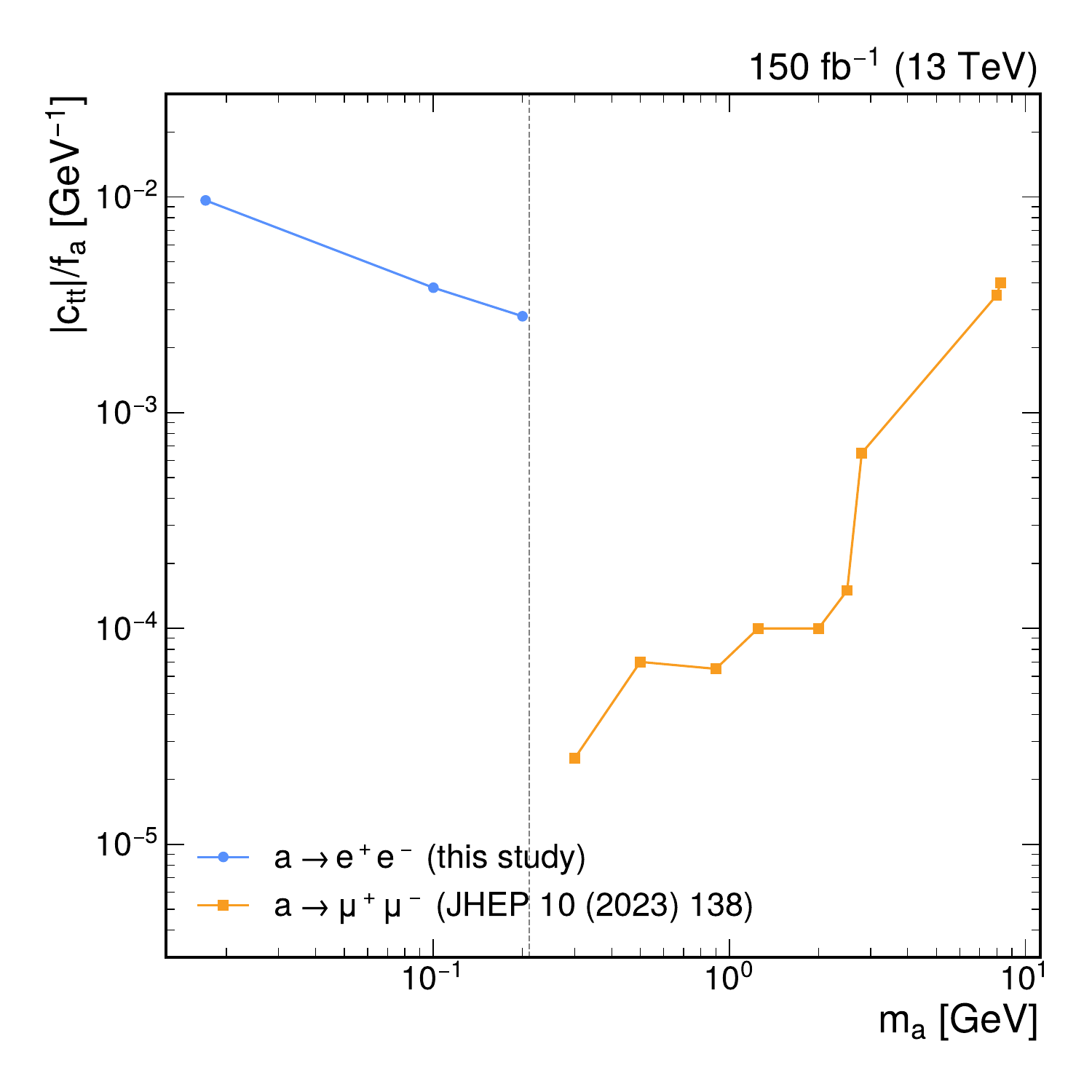}
    \caption{Expected 95\% CL upper limits on $|c_{tt}|/f_a$ as a function of the ALP mass $m_a$, in the top scenario, for an integrated luminosity of 150~fb$^{-1}$. Below the dimuon threshold (dashed vertical line), the blue circles show the smallest coupling excluded by this displaced-dielectron study, while, above the threshold, the orange squares show the expected limits of the displaced-dimuon study of Ref.~\cite{rygaard2023}. Lines connecting the mass points are shown to guide the eye.}
    \label{fig:comparison_topsecrets}
\end{figure}

For comparison with the displaced-dimuon strategy of Ref.~\cite{rygaard2023}, Fig.~\ref{fig:comparison_topsecrets} shows the expected limits on $|c_{tt}|/f_a$ for an integrated luminosity of 150~fb$^{-1}$ at $\sqrt{s} = 13$ TeV, matching the luminosity used for the Run 2 projection in Ref.~\cite{rygaard2023}. The dielectron channel extends the top-associated strategy below the dimuon threshold, $m_a = 2 m_\mu$, where the displaced-dimuon search is kinematically inaccessible. Above this threshold, the dielectron branching fraction becomes strongly suppressed in the top scenario, while the dimuon channel provides sensitivity. The comparison illustrates the complementarity of the two final states; differences in their absolute sensitivities should not be interpreted as a direct comparison of experimental performance, since the two studies use different analysis and background-modeling assumptions.

\clearpage
\pagebreak
\FloatBarrier
\justifying
\bibliographystyle{JHEP}
\justifying
\bibliography{ttalp}

\end{document}